\documentclass[twocolumn]{aastex631}

\usepackage{amsmath}
\usepackage{mathtools}
\usepackage{xspace}

\newcommand{\uas}{$\mu$as\xspace}

\newcommand{\edt}[1]{#1}

\def\grff{GRFFE\xspace} 

\accepted{March 19, 2025}

\shorttitle{Limb-Brightened Jet in M87 from Anisotropic Nonthermal Electrons}
\shortauthors{Tsunetoe et al.}

\begin{document}

\title{Limb-Brightened Jet in M87 from Anisotropic Nonthermal Electrons}

\correspondingauthor{Yuh Tsunetoe}
\email{ytsunetoe@fas.harvard.edu}

\author[0000-0003-0213-7628]{Yuh Tsunetoe}
\affiliation{Black Hole Initiative at Harvard University, 20 Garden Street, Cambridge, MA 02138, USA}
\affiliation{Center for Computational Sciences, University of Tsukuba, \ 1-1-1 Tennodai, Tsukuba, \ Ibaraki 305-8577, Japan}

\author[0000-0002-5278-9221]{Dominic~W.~Pesce}
\affiliation{Center for Astrophysics $|$ Harvard \& Smithsonian, 60 Garden Street, Cambridge, MA 02138, USA}
\affiliation{Black Hole Initiative at Harvard University, 20 Garden Street, Cambridge, MA 02138, USA}

\author[0000-0002-1919-2730]{Ramesh~Narayan}
\affiliation{Black Hole Initiative at Harvard University, 20 Garden Street, Cambridge, MA 02138, USA}
\affiliation{Center for Astrophysics $|$ Harvard \& Smithsonian, 60 Garden Street, Cambridge, MA 02138, USA}

\author[0000-0003-2966-6220]{Andrew Chael}
\affiliation{Princeton Gravity Initiative, Jadwin Hall, Princeton University, Princeton, NJ, USA}

\author[0000-0001-8053-4392]{Zachary~Gelles}
\affiliation{Department of Physics, Princeton University, Princeton, NJ 08540, USA}

\author[0000-0001-7451-8935]{Charles Gammie}
\affiliation{Astronomy Department and Physics Department, University of Illinois, 1110 West Green Street, Urbana, IL 61801, USA}

\author[0000-0001-9185-5044]{Eliot Quataert}
\affiliation{Department of Astrophysical Sciences, Princeton University, Princeton, NJ, USA}

\author[0000-0002-7179-3816]{Daniel Palumbo}
\affiliation{Black Hole Initiative at Harvard University, 20 Garden Street, Cambridge, MA 02138, USA}
\affiliation{Center for Astrophysics $|$ Harvard \& Smithsonian, 60 Garden Street, Cambridge, MA 02138, USA}



\begin{abstract}
Very long baseline interferometry observations reveal that relativistic jets like the one in M87 have a limb-brightened, double-edged structure. 
Analytic and numerical models struggle to reproduce this limb-brightening. 
We propose a model in which we invoke anisotropy in the distribution function of synchrotron-emitting nonthermal electrons such that electron velocities are preferentially directed parallel to magnetic field lines, 
as suggested by recent particle-in-cell simulations of electron acceleration and the effects of synchrotron cooling.
We assume that the energy injected into nonthermal electrons is proportional to the jet Poynting flux, and we account for synchrotron cooling via a broken power-law energy distribution. 
We implement our emission model in both 
general relativistic magnetohydrodynamic (GRMHD) simulations and axisymmetric force-free electrodynamic (\grff) jet models and produce simulated jet images at multiple scales and frequencies using polarized general relativistic radiative transfer.
We find that the synchrotron emission is concentrated parallel to the local helical magnetic field and that this feature produces limb-brightened jet images on scales ranging from tens of microarcseconds to hundreds of milliarcseconds in M87. 
We present theoretical predictions for horizon-scale M87 jet images at 230 and 345 GHz that can be tested with next generation instruments.
Due to the scale-invariance of the GRMHD and \grff models, our emission prescription can be applied to other targets and serve as a foundation for a unified description of limb-brightened synchrotron images of extragalactic jets. 

\end{abstract}

\keywords{black hole physics -- galaxies: jets -- radiative transfer -- M87 -- astroparticle physics}


\section{Introduction} \label{sec:intro}

The active galactic nucleus (AGN) Messier 87* (hereafter M87*) in the giant elliptical galaxy Messier 87 is a prime target for the study of relativistic jets from black holes (BHs). As a relatively nearby AGN ($D=16.9$\,Mpc) with an extremely massive central BH ($M_\bullet=6.2\times10^9M_\odot$; \citealt{2019ApJ...875L...5E,2021ApJ...910L..13E}),
M87* offers the opportunity of directly imaging and studying a relativistic jet from extragalactic scales down to the launching region just outside the event horizon.
In M87*, the projected gravitational radius is $\theta_{\rm g}=r_{\rm g}/D \approx 3.8\,{\mu}$as,\footnote{Throughout, we use the gravitational length scale $r_{\rm g}\equiv GM_\bullet/c^2$ and time scale $t_{\rm g}\equiv GM_\bullet/c^3$, where $M_\bullet$ is the black hole mass.} which is larger than for any other known BH except the Galactic Center BH Sagittarius A* ($\theta_{\rm g}\approx 5\,\mu$as). M87* has a powerful jet ($P_{\rm jet} \approx 10^{43}~{\rm erg\,s^{-1}}$) which is observed in all bands of the electromagnetic spectrum \citep[e.g.,][]{2002ApJ...564..683M,2006ARA&A..44..463H,2011ApJ...743..119P,2012ApJ...746..151A}. It is thus an excellent target for investigating the physics of relativistic jets and, in particular, for attempting to validate the  Blandford-Znajek process (\citealt{1977MNRAS.179..433B}) as the central acceleration mechanism in BH jets.

In this context, the successful direct imaging of the central BH in M87* by the Event Horizon Telescope (EHT; \citealp{2019ApJ...875L...1E,2019ApJ...875L...2E,2019ApJ...875L...3E,2019ApJ...875L...4E,2019ApJ...875L...5E,2019ApJ...875L...6E,2021ApJ...910L..12E,2021ApJ...910L..13E,2023ApJ...957L..20E}) is a landmark achievement that has opened a window to the physics of jet acceleration, formation, and emission on scales just outside the central supermassive black hole. 
Upcoming observations with projects like the next-generation EHT (ngEHT; \citealp{2023Galax..11..107D}) and the Black Hole Explorer (BHEX; \citealp{2024SPIE13092E..2DJ}) are expected to monitor the dynamics of the innermost regions of the jet, where the relativistic magnetized outflow emerges from the BH ergosphere. 

High resolution observations at multiple frequencies, covering a wide range of length scales from subparsec to kiloparsecs, have revealed that the M87 jet persistently shows limb-brightening. This is seen from the jet-launching region on sub-milliarcsecond (mas) scales (\citealp{2016ApJ...817..131H,2018ApJ...855..128W,2018A&A...616A.188K,2023Natur.616..686L}) all the way up to galactic scales of hundreds of mas (\citealp{2023MNRAS.526.5949N}; see also the triple-edge jet in \citealp{2017Galax...5....2H}). In transverse profiles of the intensity, the jet has a large limb-to-spine\footnote{We refer to the central region of the jet near its axis as the spine, and the region near the transverse jet boundary as the limb or the edge.} intensity ratio of $\sim 2-5$ on sub-mas (\citealp{2024arXiv240900540K}) and $\sim 2-3$ on several mas scale (\citealp{2018ApJ...855..128W}). 
Most VLBI observations of other jet sources do not resolve structure across the jet width. However, in those jets that are resolved in the transverse direction, prominent limb-brightening has been detected (e.g., 3C~84; \citealp{2018NatAs...2..472G}, Centaurus A; \citealp{2021NatAs...5.1017J}, NGC~315; \citealp{2024ApJ...973L..45P}). These observational results suggest that limb-brightening may be a universal feature of BH jets.

Meanwhile, there has been a persistent discrepancy between the observations and theoretical models. 
Theoretical studies using 
general relativistic magnetohydrodynamic (GRMHD) simulations and radiative transfer (GRRT) calculations 
indicate that a nonthermal component is required in the electron energy distribution function (eDF) to obtain detectable emission over an extended region of the jet, as observed in M87* at millimeter wavelengths (\citealp{2019A&A...632A...2D,2022NatAs...6..103C}).
However, these studies find, and we also confirm in previous work, that standard prescriptions for the nonthermal electron eDF do not reproduce the strongly-accented limb-brightened jet images seen in M87. Rather they produce single-edged or even spine-brightened jets (\citealp{2024arXiv241108116T}; see also \citealp{2022A&A...660A.107F,2024A&A...687A..88Z} for survey of nonthermal electron prescriptions). Recently, \citet{2024SciA...10N3544Y} examined the inner jet morphology with nonthermal electrons accelerated by the electric current density and cooled by synchrotron radiation. They obtained limb-brightening of the jet at around 1\,mas from the core when they considered a very rapidly spinning BH with $a_* \equiv a/M = 0.98$. 
However, models with moderate spin ($a_*=0.5$) and zero spin did not show limb-brightening.

Limb-brightened jets on large scales have also been studied with force-free electrodynamic (FFE) models (e.g., \citealp{2008MNRAS.388..551T,2009ApJ...697.1164B}). 
\citet{2018ApJ...868...82T} obtained a symmetrical double-edged jet for a BH with an extremely high spin ($a_* = 0.998$), assuming nonthermal electron injection at a location far ($80~r_{\rm g}$) from the jet axis and the BH, to avoid asymmetric emission due to the rotational motion of jet plasma. As yet, there is no unified model of BH jets which can produce limb-brightened images for a range of BH spin values and over a range of scales from near the BH horizon to galactic scales. 

In this paper, we consider a new approach to modeling the emission from jets. We assume that the nonthermal electrons in the jet have a highly {\it anisotropic} distribution, with velocities parallel to the magnetic field vastly exceeding perpendicular velocities.  
Particle-in-cell (PIC) simulations of relativistic magnetically dominated plasma turbulence show that electrons are accelerated into a nonthermal, power-law eDF with a strong anisotropy in their pitch-angle with respect to the local magnetic field (\citealp{2021PhRvL.127y5102C,2022ApJ...936L..27C}). 
\citet{2023ApJ...957..103G} implemented anisotropic thermal synchrotron emission and absorption in GRRT image calculations and explored its effect on the image. 
They surveyed three kinds of kinetic-scale instability inducing the anisotropy in the accretion disk, and pointed out that the precise choice can affect the size of the BH ring image. 

In this work, we implement anisotropy in the eDF of nonthermal electrons in the jet and study the effect of this anisotropy on the jet image. We model synchrotron emission and absorption from anisotropic power-law-distributed electrons and carry out GRRT ray-tracing to calculate theoretical M87 jet images over a range of scales using both GRMHD simulations and general relativistic FFE (\grff) models. We show that anisotropic nonthermal electrons naturally produce limb-brightened jets in a variety of scenarios.

We introduce the GRMHD and \grff models in Secions \ref{subsec:GRMHDmodel} and \ref{subsec:GRFFmodel}, our model of nonthermal electrons in Section \ref{subsec:Electrons}, and the GRRT calculations in  Section \ref{subsec:GRRT}. 
We demonstrate in Section \ref{subsec:average} that anisotropy in the eDF of nonthermal electrons is essential for producing limb-brightening in the jet image with GRMHD models. Then, in Section \ref{subsec:snapshots}, we compare the model images with existing observations and present theoretical predictions for what we might see with more sensitive and better resolved future observations. We also confirm in Section \ref{subsec:GRFF}, using the \grff model, that jet limb-brightening persists in the model up to galactic scales, as seen in observations. Finally, Section \ref{sec:discussion} is dedicated to a discussion of applications of the model and future prospects, along with concluding remarks.

\bigskip

\section{Method} \label{sec:method}

We work with two kinds of models: GRMHD models and \grff models.   GRMHD models (described in \S\ref{subsec:GRMHDmodel}) take the density, temperature, velocity, and magnetic field strength from a GRMHD simulation, and supplement these with a prescription (\S\ref{subsec:Electrons}) for nonthermal electrons in the jet. \grff models (\S\ref{subsec:GRFFmodel}) use an analytical general relativistic force-free electrodynamics model  of a time-stationary, axisymmetric jet (\citealp{2008MNRAS.388..551T,2024arXiv241000954G}), and use the same prescription (\S\ref{subsec:Electrons}) for nonthermal electrons.


The GRMHD models enable us to study individual snapshots as well as time variations. They are our preferred choice for modeling observations and studying jet dynamics close to the BH. GRMHD snapshots can also be summed to obtain time-averaged results. However, these models are reliable only out to  $r\sim 10^3r_g$, or projected angular offsets ${\sim}1$\,mas in M87. The \grff models, on the other hand, can be extended to arbitrarily large distance from the BH (we go out to ${\sim}10^5r_g$ in this paper). However, since they assume steady state, they are limited to time-averaged studies. Apart from these differences, a feature of the present work is that we use the same prescriptions to model the heating and cooling of nonthermal electrons in both the GRMHD and \grff models. Thus, in principle, for the same object we can use the GRMHD model close to the BH and switch to the \grff model at larger distances.

\subsection{GRMHD Model} \label{subsec:GRMHDmodel}

We use the same datasets as in \citet{2024arXiv241108116T}, viz., the magnetically arrested disk (MAD, \citealt{2003ApJ...592.1042I,2003PASJ...55L..69N,2011MNRAS.418L..79T,2014ARA&A..52..529Y}) GRMHD simulations reported in \citet{2022MNRAS.511.3795N}. We focus on two representative BH spin values: $a_* = +0.9$ and $+0.5$. We set \edt{the BH mass $M = 6.2\times10^9M_\odot$ \citep{2019ApJ...875L...5E,2021ApJ...910L..13E} and the mass accretion rate $\dot{M} = 5 \times 10^{-4} M_\odot {\rm yr^{-1}}$ for $a_* = +0.9$ and $\dot{M} = 1.1 \times 10^{-3} M_\odot {\rm yr^{-1}}$ for $a_* = +0.5$ (we discuss below how we estimate $\dot{M}$)}, as appropriate for M87*.
For each simulation, we consider a subset of 100 snapshots which cover a duration of $5000\,t_{\rm g}$, which corresponds to about five years in M87*. 

In each GRMHD snapshot, we divide the simulation volume into a region with low magnetization $\sigma \equiv B^2/4\pi \rho c^2 \leq 1$, and a region with high $\sigma>1$. We identify the former with the disk and the sub-relativistic wind and assume that the electrons there are purely thermal. We use 
the ``$R$-$\beta$'' prescription from \citet{2016A&A...586A..38M},
\begin{equation}
    R \equiv \frac{T_{\rm i}}{T_{\rm e}} = R_{\rm low}\frac{1}{1+\beta^2} + R_{\rm high}\frac{\beta^2}{1+\beta^2},
\end{equation}
to determine the electron temperature $T_{\rm e}$, assuming the adiabatic indices of electrons and ions to be 4/3 and 5/3, respectively. 
Here, $T_{\rm i}$ is the ion temperature and $\beta$ is the thermal-to-magnetic pressure ratio. 
We set $R_{\rm low} = 10$ and $R_{\rm high} = 160$ based on the model scoring results shown in Table 3 in \citet{2021ApJ...910L..13E} which indicate that both $a_*=0.9$ and 0.5 models with these parameters ``pass" the polarimetric and other constraints used in that study. 

We identify the $\sigma>1$ region of a snapshot with the relativistic jet. GRMHD simulations have numerical difficulties in this region because of the low gas density, which requires the application of an artificial density floor. Despite this complication, the simulations are believed to provide accurate results for quantities like the fluid velocity, magnetic field strength and Poynting flux. We therefore use the velocity and magnetic field data from the GRMHD simulation, but ignore the less-reliable density and temperature data. As an added motivation for neglecting the simulated thermal electrons, we note that the radiation from jets is known to be dominated by nonthermal emission, and nonthermal electrons are not modeled as a separate population in most GRMHD simulations (but see \citealt{2017MNRAS.470.2367C}). We estimate the electron number density and Lorentz factor distribution of the nonthermal electrons from the local Poynting flux, as described in \autoref{subsec:Electrons} and \autoref{app:eDF}.

Apart from limiting the nonthermal electrons to the region with $\sigma > 1$, we also apply an upper cutoff $\sigma_m$ so that the radiating region lies between $1<\sigma<\sigma_m$. The motivation is to limit the radiating zone to a sheath near the outer edge of the jet, based on the expectation that most of the dissipation and particle acceleration is likely to take place at this boundary. We choose
\begin{equation}\label{eq:msigmacutoff}
    \sigma_{\rm m} = \frac{300}{\sqrt{r}}.
\end{equation}
This gives a large cutoff $\sigma_m$ near the BH, covering the region where nonthermal electrons are preferentially injected by the Poynting flux (\autoref{subsec:Electrons}). 
At large radii, $\sigma_m$ becomes small (e.g., $\sim 10$ at $r = 10^3~r_{\rm g}$).

\edt{Since GRMHD simulations are scale-free, we follow the usual practice of adjusting $\dot{M}$ so as to satisfy a luminosity constraint obtained from observations. In the present work,} we set the mass accretion rate onto the BH in the $a_*=0.9$ model to $\dot{M} = 5 \times 10^{-4} M_\odot {\rm yr^{-1}}$ which gives an average flux density of ${\sim} 0.5~{\rm Jy}$ at 230~GHz for purely thermal electrons restricted to $\sigma \le 1$. This flux density matches EHT measurements of M87*. 
For the $a_*=0.5$ model, we correspondingly set $\dot{M} = 1.1 \times 10^{-3} M_\odot {\rm yr^{-1}}$.
Both of these $\dot{M}$ values are in good agreement with the corresponding estimates \edt{($8 \times 10^{-4} M_\odot {\rm yr^{-1}}$, $1.6 \times 10^{-3} M_\odot {\rm yr^{-1}}$, respectively)} given in Table 3 in \citet{2021ApJ...910L..13E}. \edt{The small differences are consistent with the natural variability we expect in simulations, given that the \citet{2021ApJ...910L..13E} work was based on an earlier set of GRMHD simulations run with the \texttt{iharm} code \citep{Gammie2003,noble2006primitive}, whereas our work is based on the simulations from \cite{2022MNRAS.511.3795N} which used different initial conditions and were run with the \texttt{KORAL} code \citep{2013MNRAS.429.3533S,2014MNRAS.439..503S}.}  

In \autoref{sec:results}, we show images corresponding to the $a_* = +0.9$ GRMHD simulation, and treat this as our fiducial model. We discuss the second model with $a_* = +0.5$ in \autoref{sec:discussion}.

\subsection{\grff Model} \label{subsec:GRFFmodel}

To describe the properties of the jet on scales larger than those accessible to GRMHD simulations, we employ an axisymmetric \grff model in which the fluid inertia and internal energy are negligible and only electromagnetic forces are dynamically relevant.  We use the \grff jet model developed by \citet{2024arXiv241000954G}, which builds on earlier work by \citet{2008MNRAS.388..551T} and \citet{2009ApJ...697.1164B} and extends the model to a Kerr spacetime.  This model provides a description of the jet geometry, magnetic field structure, and bulk fluid Lorentz factor throughout the jet.

The magnetic field structure in the \grff model is determined by the so-called ``magnetic stream function'' $\psi$, which is related to the azimuthal component of the vector potential by $\psi = A_{\phi}$.\footnote{This expression for $\psi$ is appropriate for the Boyer-Lindquist coordinates used in the \citet{2024arXiv241000954G} GR model, though we note that some previous papers \citep[e.g.,][]{2007MNRAS.375..548N,2008MNRAS.388..551T} used orthonormal coordinates in flat space such that in those coordinates $\psi = R A_{\phi}$.}
For an axisymmetric system, $\psi$ must satisfy a second-order differential equation \citep[see][]{1974MNRAS.167..457O,2007MNRAS.375..548N} that admits the approximate solution
\begin{equation}
\psi(r,\theta) \propto \begin{dcases}
r^{2-2s} \Big[ 1 - \cos(\theta) \Big] , & 0 \leq \theta \leq \frac{\pi}{2} \\
r^{2-2s} \Big[ 1 + \cos(\theta) \Big] , & \frac{\pi}{2} < \theta \leq \pi
\end{dcases} , \label{eqn:StreamFunction}
\end{equation}
\noindent which is accurate to $\lesssim$10\% for both non-rotating and rotating systems in the flat space limit \citep{2008MNRAS.388..551T}.  The specification of $2-2s$ for the radial power-law index ensures that at large $r$ the magnetic field lines follow a jet shape described by
\begin{equation}
R \propto z^s , \label{eqn:CollimationRate}
\end{equation}
\noindent such that $s = 0.5$ corresponds to a parabolic jet and $s = 1$ corresponds to a conical jet.  For this paper, we adopt a value of $s = 0.6$, which is consistent with VLBI measurements of the M87 jet \citep[e.g.,][]{2013ApJ...775...70H,2016ApJ...817..131H,2018ApJ...868..146N}. The magnetic field is determined from $\psi$ following Eq.\,6 in \citet{2024arXiv241000954G}. All magnetic field lines in the \grff model live on contours of constant $\psi$, and the field lines that intersect the black hole horizon at poloidal angle $\theta = \pi/2$ define the edge of the jet.

To set the overall normalization of the magnetic field, we assume that the total jet power -- measured from the integrated Poynting flux in both jets -- is equal to
\begin{equation}
P_J = 1.4 a_*^2 \dot{M} c^2 , \label{eqn:JetPower}
\end{equation}
where $\dot{M}$ is the mass accretion rate onto the black hole. The coefficient 1.4 is appropriate for accretion in the magnetically arrested disk (MAD) state \citep{2011MNRAS.418L..79T,2022MNRAS.511.3795N}. 
We use the same value of $\dot{M}$ as in the GRMHD model. Also, we assume that the jet radiation is produced by nonthermal electrons, which we model using the local Poynting flux, as in the GRMHD model. Details are given in Section~\ref{subsec:Electrons}.

The magnetic field lines rotate rigidly \citep{1937MNRAS..97..458F} at a rate determined by the black hole's spin and the location at which the field lines thread the horizon \citep[e.g.,][]{1977MNRAS.179..433B}.  The fluid motion perpendicular to magnetic field lines is set by the force-free condition to be the so-called ``drift velocity,'' which can be analytically determined from the known $E$- and $B$-fields \citep[e.g.,][]{2006MNRAS.368.1561M,2007MNRAS.375..548N,2023ApJ...958...65C}.  However, the force-free condition does not by itself constrain the component of the fluid motion parallel to the magnetic field lines, which has thus often been assumed to be zero \citep[e.g.,][]{2008MNRAS.388..551T,2009ApJ...697.1164B}.  For our \grff model, we instead use the prescription from \citet{2024arXiv241000954G} that assigns a value for the parallel fluid motion using an energy conservation argument in the force-free limit of cold GRMHD; see \citet{2024arXiv241000954G} for details.

Unlike in GRMHD, the Lorentz factor of plasma bulk motion $\gamma_{\rm bulk}$ in the \grff model can grow very large ($\gamma_{\rm bulk} \gg 1$) far away from the BH, reaching $\sim 20 - 80$ at $z \sim 10^4 - 10^5 r_{\rm g}$.
To allow for drag on the jet at its boundaries, we suppress $\gamma_{\rm bulk}\beta_{\rm bulk}$ by a factor of 0.5\footnote{\edt{The factor of 0.5 is a rough estimate, which we obtained by comparing the velocity profile in the GRMHD simulation with the corresponding velocity in the \grff model.}} throughout the entire \grff jet, where $\beta_{\rm bulk}$ is the bulk velocity. 
In addition, we set a hard ceiling of $\gamma_{\rm bulk} \le 6$.\footnote{Note that this hard ceiling differs from the prescription used in \citet{2024arXiv241000954G}, who instead apply a softer cutoff (see their Eq. 49).}
Together, the velocity suppression and ceiling produce fluid Lorentz factors in the GRFFE model that are closer to those found in the GRMHD simulations. 
Furthermore, we confirm in Section \ref{subsec:GRFF} that the suppression and ceiling are favored in comparison with observed images of the M87 jet, which serves as a validation of the prescription here.

In analogy with the sigma cutoff $\sigma_m$ applied in the GRMHD model, we exclude the spine region in the \grff model as well. 
We cut out the region with 
\begin{equation}
    \psi(r,\theta) < \psi(r_{\rm fp},\theta_{\rm fp})
\end{equation}
for the northern jet, and the symmetrical region in the southern hemisphere; here $r_{\rm fp} = r_{\rm H}$, the horizon radius,  and $\theta_{\rm fp}$ is set to $65^\circ$. This corresponds to a cutoff of  the inner $\sim 80~\%$ of the jet width at large distances.

\subsection{Nonthermal Electrons in the Jet} \label{subsec:Electrons}

The GRMHD and \grff models describe the structure of the magnetic field and fluid velocity throughout the jet, but they need to be supplemented with a model for the number density and energy distribution of synchrotron-emitting electrons. There is strong observational evidence that the radiation we observe from jets is produced primarily by nonthermal (power-law) electrons and that these electrons are continuously accelerated throughout the length of the jet, probably through dissipative interactions with the surrounding medium \citep[e.g.,][]{1984ARA&A..22..319B}. With this in mind, we assume the following prescriptions for the heating and subsequent synchrotron cooling of nonthermal electrons (\autoref{app:eDF} discusses additional details).

In the interest of simplicity, we make a number of approximations. We assume that the energy to heat the electrons in the jet comes from the jet power, for which  we use the
\edt{Poynting flux $\vec{S}$ in the Zero-Angular-Momentum Observer (ZAMO) frame as a proxy (see \autoref{apdx:poynting} for details).}
This quantity is readily available in both the GRMHD and \grff models.
The dissipation of jet power is likely related to ``frictional" drag exerted on the relativistic jet by instabilities where it meets the external medium (which is either a slowly moving wind, or a stationary ambient medium); the actual mechanism of particle acceleration may be related to magnetic reconnection in the shear flow that will naturally develop at the jet boundary \citep[e.g.,][]{2021ApJ...907L..44S}. In this picture, we expect the bulk of the heating to occur in a relatively narrow region near the outer boundary of the jet. In the force-free region of a jet, the Poynting flux itself is largest at the outer edge and thus naturally introduces preferential edge-heating if we simply make the heating rate proportional to the Poynting flux. In addition, as explained in Sections \ref{subsec:GRMHDmodel} and \ref{subsec:GRFFmodel}, we accentuate the edge-heating effect by arbitrarily cutting out the central (``spine") region of the jet and setting the heating in this interior region to zero. 

We adopt here the ansatz that some small fraction $h \ll 1$ of the jet power is converted per logarithmic interval of $r$ into nonthermal electron energy density.  In this spirit, we write the accumulated nonthermal electron energy density injected into a given jet fluid element as
\begin{equation}
    u_{\rm nt,inj} = h\,\frac{|\vec{S}|}{c} . \label{eq:uplinj}
\end{equation}
To motivate this formula, the electromagnetic energy density associated with the Poynting flux is $|\vec{S}|/c$, and a fraction $h$ of this energy density is injected in the form of nonthermal electrons.  

We assume that the injected electrons have a power-law distribution of Lorentz factor $\gamma$ of the form
\begin{eqnarray}\label{eq:singlePL}
   f_{\rm inj}(\gamma) = \begin{dcases}
0 , & \gamma < \gamma_m \\
\frac{n_{\rm pl}(p-1)}{\gamma_m ^{1-p}-\gamma_{\rm max}^{1-p}}\gamma^{-p}, & \gamma_m \leq \gamma \leq \gamma_{\rm max} \\ 
 0, & \gamma > \gamma_{\text{max}}
\end{dcases} .
\label{eq:ninj}
\end{eqnarray}
\edt{Here, $\gamma_m$ and $\gamma_{\rm max}$ are the minimum and maximum Lorentz factors of the injected electrons.} Equating the total energy in these electrons (including the rest mass energy, though the latter is small since the $\gamma$ values we consider are large) to $u_{\rm nt,inj}$,
we determine the power-law electron number density $n_{\rm pl}$ to be
\begin{equation}
    n_{\rm pl} = \frac{u_{\rm nt,inj}(p-2)(\gamma_m^{1-p} - \gamma_{\text{max}}^{1-p})}{m_{\rm e} c^2(p-1)(\gamma_m^{2-p} - \gamma_{\text{max}}^{2-p})}.
\end{equation}
The actual $f(\gamma)$ at any location in the jet is different from $f_{\rm inj}(\gamma)$ because of synchrotron cooling. The oldest injected electrons have cooled for a time of order the characteristic dynamical time (assuming the jet travels at a speed $\sim c$),
\begin{equation}
t_c = \frac{r}{c} , \label{eqn:CharacteristicTimescale}
\end{equation}
\noindent which corresponds to a synchrotron cooling Lorentz factor (see Equations~\ref{Eq:gammadot}, \ref{Eq:Gamma} in \autoref{app:eDF}, and the associated caveat in a footnote)
\begin{equation}
\gamma_c \equiv \Gamma(t_c) = \frac{6 \pi m_e c}{\sigma_T B^2 t_c} . \label{eqn:GammaC}
\end{equation}
\noindent Electrons with injected Lorentz factors $\gamma > \gamma_c$ are able to cool to $\gamma_c$ over a time $t_c$. For those electrons injected more recently and hence have ages $t<t_c$, the corresponding $\Gamma(t)$ is larger. The net $n(\gamma)$ is obtained by integrating over all ages between 0 and $t_c$. This is a standard problem in synchrotron theory \citep[e.g.,][]{1998ApJ...497L..17S} and results in a double power-law distribution of electron Lorentz factors.

Following \citet{1998ApJ...497L..17S}, we 
distinguish between two cooling regimes: a ``slow-cooling" regime that corresponds to $\gamma_c \geq \gamma_m$, and a ``fast-cooling" regime that corresponds to $\gamma_c < \gamma_m$. The corresponding electron energy distributions, $n_{\rm slow}(\gamma)$ and $n_{\rm fast}(\gamma)$, are given by Equations \eqref{eqn:SlowCoolingApprox} and \eqref{eqn:FastCoolingApprox} in \autoref{app:eDF}.

The above model for the nonthermal electron energy distribution has four free parameters -- $h$, $p$, $\gamma_m$, and $\gamma_{\rm max}$ -- where we may safely set $\gamma_{\rm max}\to\infty$ in most situations\footnote{\edt{We cannot set $\gamma_{\rm max}\to\infty$ if $p\leq2$, since the energy content in the injected electrons would diverge. For such values of $p$, $\gamma_{\rm max}$ should be treated as an important adjustable parameter. However, since in the present work we choose $p=2.5$, we may set $\gamma_{\rm max}$ to any large value we please; we arbitrarily choose $\gamma_{\rm max}=10^8$.}}. We use the same parameter values for both the GRMHD and \grff models, simply replacing the Poynting flux estimate appropriately for each model. We recognize that the present model is fairly unsophisticated, ignoring time dilation and other relativistic effects, but we feel that it is adequate for an initial trial model such as the present study.\footnote{All special and general relativistic effects are treated correctly when we do ray-tracing to calculate images. The only exception is that the ray-tracing code works in the ``fast light" approximation (see \autoref{subsec:GRRT}).}  The current model also assumes that the synchrotron cooling takes place in a constant-strength magnetic field, rather than one whose magnitude decreases with distance from the black hole (see \autoref{app:eDF}).

For this paper, we adopt $p = 2.5$ \edt{(a fairly standard value)} for the power-law index and $\gamma_m=30, ~\gamma_{\rm max}=10^8$ for the minimum and maximum Lorentz factors of the injected electron distribution\footnote{\edt{The choice $\gamma_m=30$ is informed by the following argument. The hot two-temperature accretion flow from which M87's jet originates has an electron temperature $T_e \sim {\rm few}\times 10^{10}$\,K, for which the thermal Maxwell-J\"uttner distribution function peaks at a Lorentz factor $\gamma_{\rm peak}\sim10$. Assuming that the power-law distribution of the accelerated electrons in the jet begins a factor of a few above $\gamma_{\rm peak}$, we choose $\gamma_m = 30$.}}.  We set $h=0.0025$ for the fiducial model ($a_* = 0.9$) which gives for the GRMHD model an average total flux density (thermal and nonthermal radiation) at 86~GHz of ${\sim} 1.1~{\rm Jy}$, in agreement with observations. In the secondary model ($a_* = 0.5$), we similarly find $h = 0.005$. 

\subsection{GRRT with Anisotropic Nonthermal Electrons} \label{subsec:GRRT}

As mentioned in \S\ref{sec:intro}, a key feature of the present work is the inclusion of an anisotropic angular energy distribution for the nonthermal electrons in the jet. Here, we derive the synchrotron emission and absorption coefficients for an anisotropic power-law distribution of Lorentz factor $\gamma$, in an analogous way with the thermal case considered in \citet{2023ApJ...957..103G}. We consider here a single power-law \edt{energy distribution function} and then apply the same analysis to the double power-law case in \autoref{apdx:aniso_wPL}.

Following \citet{2023ApJ...957..103G}, we write the anisotropic single power-law distribution function as 
\begin{eqnarray} \label{eq:ano_def}
    f(\gamma,\xi) &=& N_{\rm pl} \left\{ \sqrt{1 + \left(\frac{p_\perp}{m_{\rm e} c}\right)^2 + \eta\left(\frac{p_\parallel}{m_{\rm e} c}\right)^2} \right\}^{-p} \nonumber \\
    &=& N_{\rm pl} \left\{ \sqrt{1 + (\gamma^2-1)({\rm sin}^2\xi + \eta \ {\rm cos}^2\xi)} \right\}^{-p} \nonumber \\
    && (\gamma_m < \gamma < \gamma_{\rm max}, ~0 \leq \xi \leq \pi).
\end{eqnarray}
Here, $\gamma$ and $\xi$ are the Lorentz factor of an electron and the pitch angle of the electron's motion with respect to the magnetic field; 
$p_\perp = m_{\rm e} c\sqrt{\gamma^2-1}\,{\rm sin}\,\xi$ and $p_\parallel = m_{\rm e} c\sqrt{\gamma^2-1}\,{\rm cos}\,\xi$ are the relativistic momentum components perpendicular and parallel to the magnetic field; 
$N_{\rm pl}$ is a normalization factor with units of $[{\rm cm}^{-3}]$; $p$ is the power-law index; and $\gamma_m$ and $\gamma_{\rm max}$ are the minimum and maximum Lorentz factors. 
The parameter $\eta$ describes the degree and sense of anisotropy; 
$\eta < 1$ (or $> 1$) indicates that the motion of electrons is concentrated along (perpendicular to) the direction of the local magnetic field; 
$\eta = 1$ corresponds to the usual isotropic power-law distribution. 

For large $\gamma$, 
\begin{equation} \label{eq:anisopl}
    f(\gamma,\xi) = N_{\rm pl} \ \gamma^{-p} \left\{1 + (\eta-1){\rm cos}^2\xi \right\}^{-p/2} .
\end{equation}
Normalization by the total number density of power-law electrons $n_{\rm pl}$ gives 
\begin{equation}
    N_{\rm pl} = \frac{(p-1)n_{\rm pl}}{\gamma_m ^{1-p}-\gamma_{\rm max}^{1-p}} P(p,\eta)^{-1},
\end{equation}
\begin{eqnarray}
    P(p,\eta) &=& \frac{1}{2}\int_0^{\pi} {\rm d}\xi \ {\rm sin}\xi \left\{1 + (\eta-1){\rm cos}^2\xi \right\}^{-p/2} \\
        &=& \int_0^1 {\rm d}\mu \ \left\{1 + (\eta-1)\mu^2 \right\}^{-p/2}.
\end{eqnarray}
Thus, the anisotropic single power-law distribution function can be written as 
\begin{equation} \label{eq:anisopl_fin}
    f(\gamma,\xi) = \phi(\xi) \ f_{\rm iso}(\gamma),
\end{equation}
where
\begin{equation}
    \phi(\xi) = P(p,\eta)^{-1} \left\{1 + (\eta-1){\rm cos}^2\xi \right\}^{-p/2},
\end{equation}
and $f_{\rm iso}(\gamma)$ corresponds to the isotropic single power-law distribution function, i.e., \autoref{eq:anisopl} with $\eta=1$.
\autoref{eq:anisopl_fin} takes a simpler form than the thermal case discussed in \citet{2023ApJ...957..103G} in that it separates into the product of $\xi$- and $\gamma$-dependent parts, as assumed in \citet{1971Ap&SS..12..172M}. 

The anisotropic power-law emissivity is obtained by integrating the single particle emissivity with the distribution function in Eq.~\ref{eq:anisopl_fin}. 
For the emitting electrons with large Lorentz factor, the synchrotron radiation is strongly beamed in the direction of the pitch angle $\xi \simeq \theta_B$, where $\theta_B$ is the angle between the light ray and the magnetic field. 
\citet{1971Ap&SS..12..172M} wrote down the emissivity and absorption coefficients expanded to the first order of $\gamma^{-1}$ around $\xi = \theta_B$. 
In our case with Eq.~\ref{eq:anisopl_fin}, the full-Stokes synchrotron emissivities and absorption coefficients are obtained as\footnote{While the total and linear polarization emissivities and absorption coefficients given here are applicable in general, the expressions for circular polarization were obtained in the limit $\gamma_{\rm m} \rightarrow 0$, $\gamma_{\rm max} \rightarrow \infty$, as in \citet{1971Ap&SS..12..172M}, and are valid only in the frequency range $\nu_m \ll \nu \ll \nu_{\rm max}$. Although this approximate expression is sufficient for our present purposes since this work focuses only on total intensity images, it is fairly straightforward to do a more accurate calculation for futurel applications by applying the correct limits in the double integral in $j_{V,{\rm iso}}$ and tabulating the results numerically.}  
\begin{eqnarray}
    j_I &=& \ \phi(\theta_B) \ j_{I,{\rm iso}}, \\
    j_Q &=& \ \phi(\theta_B) \ j_{Q,{\rm iso}}, \\
    j_U &=& \ 0, \\
    j_V &=& \ \phi(\theta_B) \ \left\{ 1 + \frac{g(\theta_B)}{p+2} \right\} \ j_{V,{\rm iso}} ~,
\end{eqnarray}
and 
\begin{eqnarray}
    \alpha_I &=& \ \phi(\theta_B) \ \alpha_{I,{\rm iso}}, \\
    \alpha_Q &=& \ \phi(\theta_B) \ \alpha_{Q,{\rm iso}}, \\
    \alpha_U &=& \ 0, \\
    \alpha_V &=& \ \phi(\theta_B) \ \left\{ 1 + \frac{g(\theta_B)}{p+2} \right\} \ \alpha_{V,{\rm iso}} \,,
\end{eqnarray}
where
\begin{eqnarray}
    g(\theta_B) &=& {\rm tan}\,\theta_B \frac{1}{\phi(\theta_B)} \left.\frac{{\rm d}\phi}{{\rm d} \xi}\right|_{\xi=\theta_B} \\
    &=& \frac{p(\eta-1){\rm sin}^2\theta_B}{1 + (\eta-1){\rm cos}^2\theta_B} \,.
\end{eqnarray}
Here, $j_{\{I,Q,V\},{\rm iso}}$ and $\alpha_{\{I,Q,V\},{\rm iso}}$ are the emissivities and absorption coefficients for the isotropic power-law case with minimum and maximum Lorentz factors $\gamma_m$ and $\gamma_{\rm max}$ (see, for example, \citealp{2016MNRAS.462..115D}, Appendix A2). 
In Appendix \ref{apdx:aniso_wPL}, we apply the above analysis to a broken, double power-law distribution of electrons. 

We have implemented the full-polarization, anisotropic double power-law synchrotron emissivity and absorption coefficients in the GRRT code \texttt{SHAKO} (see Appendix of \citealp{2024PASJ..tmp...84T} for validation of the code), to calculate multi-scale jet images based on the GRMHD and \grff models. 
We numerically tabulate the integrals in $j_{\{I,Q,V\}, {\rm iso}}$ and $\alpha_{\{I,Q,V\}, {\rm iso}}$, as in \citet{2024PASJ..tmp...84T,2024arXiv241108116T}. For the Faraday rotation and conversion coefficients, we adopt the isotropic single power-law forms.  

In this work, we assume that the nonthermal electrons are injected with $p=2.5$, $\gamma_m=30$, $\gamma_{\rm max} = 10^8$, and have a highly anisotropic distribution with $\eta=0.01$, while the thermal electrons are isotropic. \edt{A distribution with $\eta=0.01$ corresponds to an average pitch angle of $\langle{\rm sin}^2\,\xi\rangle \sim 0.2$ for $p=2.5$, $\langle{\rm sin}^2\,\xi\rangle \sim 0.1$ for $p=3.5$ (relevant above the cooling break, see \autoref{eqn:SlowCoolingApprox}), and $\langle{\rm sin}^2\,\xi\rangle \sim 0.3$ for $p=2$ (below the break in the fast-cooling regime, see \autoref{eqn:FastCoolingApprox}).} 
We set the observer's inclination angle with respect to the BH spin axis to $i = 163^\circ$\citep{2018ApJ...855..128W}. 

We note that our GRRT calculations are done under the usual ``fast light" approximation, in which one ignores propagation time differences between rays emerging from different cells in the GRMHD simulation volume. The fast light approximation is equivalent to taking the speed of light to infinity. The more proper ``slow light" approach, which uses the correct finite light speed, is computationally more involved since the image observed at a given instant receives contributions from radiation emitted at different times, and hence from different snapshot outputs of the GRMHD simulation. For EHT applications, slow light calculations are considered to be unnecessary since tests indicate that, at least for certain observables, slow light introduces only small changes in the results relative to fast light calculations \citep{2010ApJ...717.1092D,2018A&A...613A...2B}. However, for the study of beamed relativistic jets, as in the present work, the differences will be large and it is essential to include slow light in future work.


\subsection{Expected Anisotropy in the Nonthermal Electron Population} \label{subsec:anisotropy}



In strongly magnetized low collisionality plasmas such as those found in jets, there are strong theoretical and empirical reasons to expect the plasma distribution function to be anisotropic with respect to the magnetic field.   Empirically, such anisotropy is well-established in the solar corona and solar wind \citep{2009PhRvL.103u1101B}.   The most extreme anisotropy is inferred via spectroscopy of coronal holes, which implies that oxygen ions have perpendicular temperatures $\sim 10-100$ times larger than their parallel temperatures \citep{1999ApJ...511..481C}.  Note that this is the opposite sense of anisotropy as that invoked in the present paper, but there is no reason to expect oxygen heating in the solar corona and relativistic electron acceleration (and synchrotron cooling) in jets to produce the same anisotropy.  The key point of this example is that very large anisotropies are observed in the strongly magnetized solar corona.   As we now describe, several processes can generically generate anisotropy in the electron distribution function in jets.   By contrast, there are few (if any)  mechanisms known to robustly isotropize the plasma distribution function, precisely because the plasma is highly magnetized and low-collisionality.

Absent heating and cooling, adiabatic evolution of the distribution function will drive plasma anisotropy in a low-collisionality plasma. The adiabatic invariants for charged particle motion parallel ($j_\parallel$) and perpendicular ($j_\perp$) to the field are conserved for slow variations of the background state \citep[e.g.][]{Sturrock:1994}.  Following the Appendix in \cite{Chandra:2015}, $j_\perp = p_\perp^2/\tilde{B}$ and $j^\parallel = (\tilde{B}/\tilde{\rho}) p_\parallel$. Here $p_\perp, \,p_\parallel$ are the momenta perpendicular and parallel to the field, \edt{and $\tilde{B}$ and $\tilde{\rho}$ are the ratio of magnetic field strength and density to their values in some reference state.}  As density and magnetic field strength evolve, conservation of $j_\parallel, \,j_\perp$ drives anisotropy.

For example, suppose $dn_e/d^3p \propto \gamma^{-p-2}$.  Initially, in the ultrarelativistic limit,  $\gamma^{-p - 2} \sim (j_\perp + j_\parallel^2)^{-(p - 2)/2}$. Since $j_\perp, j_\parallel$ are constant under slow evolution of the field strength and density, at later times $\gamma^{2} \propto (p_\perp/\tilde{B} + p_\parallel^2 (\tilde{B}/\tilde{\rho})^2)$.  On  comparing to Equation \ref{eq:ano_def}, we see that this implies the anisotropy parameter $\eta = \tilde{B}^3/\tilde{\rho}^2$.  

The scaling factors $\tilde{B}$ and $\tilde{\rho}$ depend on jet shape and velocity. Suppose the jet is axisymmetric and has a purely toroidal magnetic field.  If in cylindrical coordinates each streamline has $R \propto z^s$ and $v_z \propto z^a$, then one can show that $\tilde{\rho} = (z/z_0)^{-2 s - a}$ and $\tilde{B} = (z/z_0)^{-s - a}$, so that $\eta = \tilde{B}^3/\tilde{\rho}^2 = (z/z_0)^{s - a}$. Then for $s = 0.6$ in an unaccelerated ($a = 0$) jet $\eta$ {\em increases} downstream.  If acceleration  is constant ($a = 1$), however, $\eta$ {\em decreases} downstream.  Even then anisotropy driving is weak: for an initially isotropic distribution at $z_0 = r_g$, $\eta$ is only $0.15$ at $z = 100 r_g$.  Although adiabatic driving of anisotropy is interesting because it links the electron anisotropy to the shape and velocity profile of the jet, it is likely too weak to produce the $\eta \sim 0.01$ assumed in our model.

Even if adiabatic evolution of the plasma can generate the needed anisotropy, another argument against it as the dominant source of anisotropy is that significant heating is needed to explain the surface brightness profiles of jets and the flat spectrum in the optically thick regime (e.g., \citealt{1979ApJ...232...34B}).

Synchrotron cooling will also generate anisotropy, as described in \cite{Zhdankin:2023}.  The radiation reaction force preferentially reduces momentum perpendicular to the magnetic field, driving the plasma toward $\eta<1$. For cooling to be effective, the cooling time must be shorter than the outflow time for electrons that produce synchrotron radiation at the frequencies of interest.  In our GRMHD model for M87 the cooling time for electrons that produce $86$~GHz emission and the dynamical time are equal at $r \sim 100 r_g$. At higher frequency and larger radius cooling can be effective in driving anisotropy.


For a model in which particle injection by acceleration is balanced by cooling, the steady state anisotropic distribution function for isotropic injection and (anisotropic) synchrotron cooling is $\propto (\sin^2 \alpha)^{-1} \gamma^{-p-1}$ where $p$  is the injection spectrum and we have assumed $p > 1$ (as is typically the case in most particle acceleration scenarios).  This follows from the fact that the steady state distribution function is proportional to the injection spectrum times the synchrotron cooling time: this steepens the spectrum by one power of $\gamma$ and introduces the $1/\sin^2 \alpha$ dependence.   Note that this synchrotron-cooled steady state distribution function is precisely of the form of equation \ref{eq:anisopl_fin} with $\eta = 0$ except that strong synchrotron cooling fixes the $\sin^2 \alpha$ dependence independent of the momentum power-law index.  

A third source of electron anisotropy is the electron acceleration process itself.  Particle-in-cell simulations of particle energization in strongly magnetized turbulent plasmas show that the electrons are preferentially accelerated along the local magnetic field, creating a pitch-angle-anisotropy with $T_\parallel > T_\perp$ \citep{2022ApJ...936L..27C}.  The same occurs in reconnection simulations with a significant mean magnetic field \citep{2023ApJ...959..137C}.  This effect is, however, less prominent at higher electron energies.

Pitch angle scattering tends to restore isotropy in the distribution function.  Both Coulomb scattering and wave-particle scattering in principle contribute.  Isotropization will be effective when the scattering timescale is shorter than the anisotropy driving timescale. In jets, the densities are low and the electron energies are high, so Coulomb scattering can be safely neglected.   Wave particle scattering can in principle be produced by the ambient turbulent fluctuations.   The properties of this scattering in relativistic plasma turbulence are not well-characterized.   In the empirically well-studied case of the solar wind, however, wave-particle scattering appears to primarily be due to kinetic instabilities generated by large anisotropy in the distribution function \citep{2009PhRvL.103u1101B}, not  fluctuations associated with the ambient turbulence.
For example, the fluid firehose instability sets in if $P_\parallel - P_\perp \gtrsim B^2/4 \pi$ (both non-relativistically and relativistically; e.g., \citealt{2023ApJ...957..103G}).   In our models $\beta \sim 10^{-2}$, permitting large anisotropies in both ions and electrons.   
Driving by radiative cooling will, however, produce anisotropy in the electrons only, and particle acceleration may preferentially do so as well.  The relevant instability is then a resonant electron-only version of the firehose instability.  This has been studied in non-relativistic electron-ion plasmas (e.g., \citealt{2008JGRA..113.7107C}) but not, to the best of our knowledge, in relativistic electron-ion or pair plasmas.  Nonetheless on energetic grounds -- and by analogy to other such kinetic instabilities -- we are confident that relatively large electron $T_\parallel/T_\perp$ will be kinetically stable in strongly magnetized jets.

\section{Results} \label{sec:results}

\subsection{Time-Averaged Images with the GRMHD Model} \label{subsec:average}

\begin{figure}
\begin{center}
	\includegraphics[width=9cm]{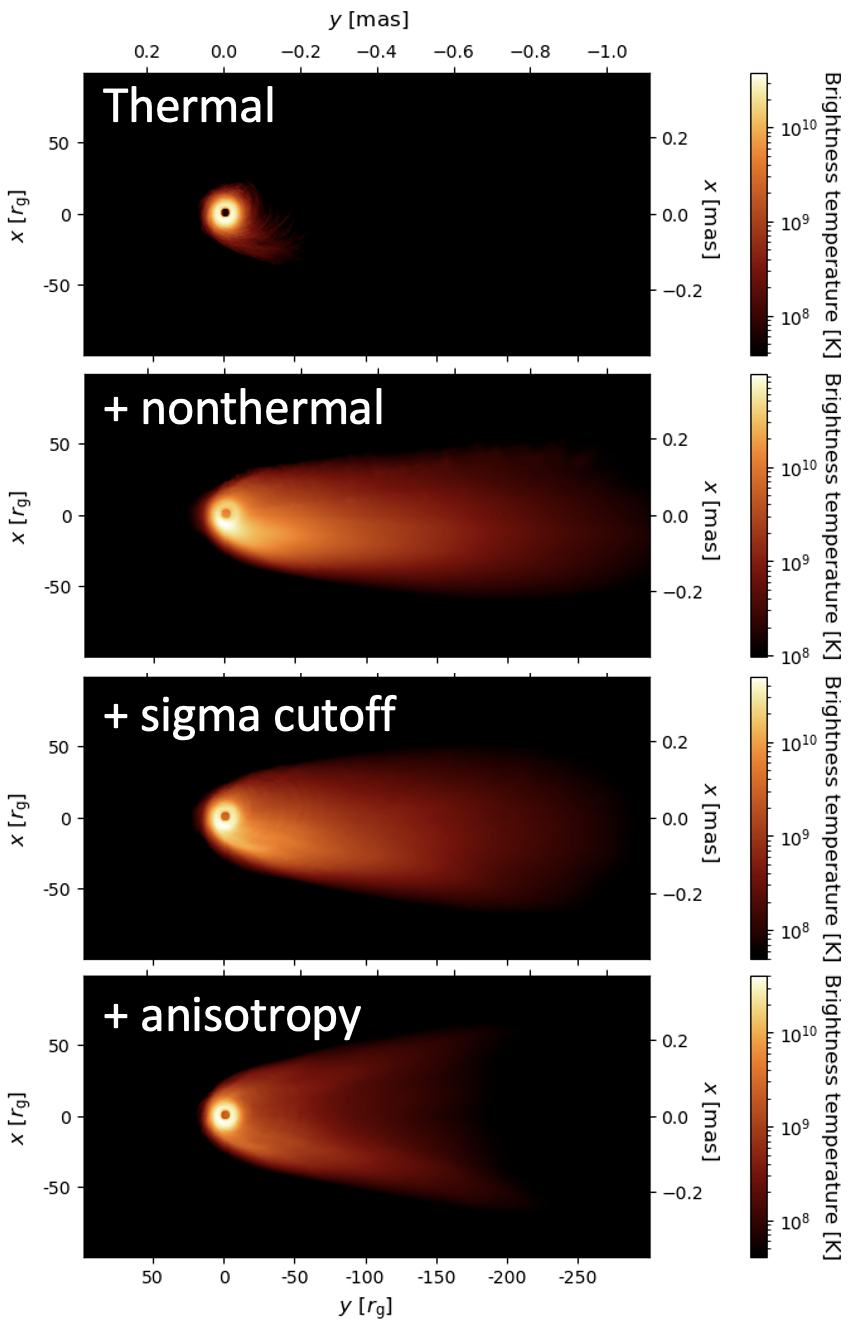}
\end{center}
    \caption{
    Time-averaged images of the M87* BH and its jet at 86~GHz, based on the $a_*=0.9$ GRMHD model, for four electron energy distribution prescriptions. 
    Top panel: assuumes thermal electrons in the disk and wind, which are defined as the region with magnetization  $\sigma \leq 1$, and no radiation from the jet, which is defined as the region with $\sigma > 1$. 
    Second panel: thermal electrons in the disk and wind, and  nonthermal electrons in the jet, modeled as described in Section \ref{subsec:Electrons}. Third panel: thermal electrons in the disk and wind, and nonthermal electrons in the jet, but with the jet emission limited to a sheath by means of a sigma cutoff, $1 < \sigma < 300/\sqrt{r}$.
    Bottom panel: Same as the third panel, but introducing a strong anisotropy, with $\eta=0.01$, in the energy distribution of the nonthermal electrons in the jet. This is the only model that produces a limb-brightened jet image similar to what is seen in time-averaged observations (see bottom right panel in \autoref{fig:GRFF_obs}).
    }
    \label{fig:thPcutaniso}
\end{figure}

First, we demonstrate that anisotropy in nonthermal electrons is an essential ingredient in our GRMHD model for producing limb-brightened images in the jet-launching region in M87. The images shown below are based on our fiducial model with $a_*=0.9$ (see section \ref{sec:discussion} for the $a_* = 0.5$ model). 

In \autoref{fig:thPcutaniso}, we show time-averaged GRMHD images over a duration of $5000~t_{\rm g}$ at 86 GHz made using purely thermal electrons in the disk and wind (which are defined as the region with magnetization $\sigma\leq 1$) and no radiation from the jet (top panel); thermal disk-wind and isotropic ($\eta=1$) nonthermal electrons in the jet region (defined as $\sigma > 1$, second panel); thermal disk-wind and isotropic nonthermal electrons in the jet with a sigma cutoff following \autoref{eq:msigmacutoff} to enhance limb-brightening (third panel); and thermal disk-wind and anisotropic ($\eta=0.01$)  nonthermal electrons in the jet with the sigma cutoff of \autoref{eq:msigmacutoff} (bottom panel). 
As explained earlier, we set $M_\bullet = 6.2\times10^9M_\odot$, $\dot{M} = 5 \times 10^{-4} M_\odot {\rm yr^{-1}}$, and choose the observer's inclination angle to be $i= 163^\circ$ \citep{2018ApJ...855..128W}.

We can clearly see in the top panel in \autoref{fig:thPcutaniso} that thermal synchrotron emission from the disk and wind produces a ring-like structure similar to that observed by the EHT at 230\,GHz, but it does not produce much extended emission along the jet even though the $\sigma\leq1$ wind extends out to a large distance. 

Injection of nonthermal, power-law electrons in the jet ($\sigma>1$) using the Poynting flux prescription of \autoref{subsec:Electrons} with $h=0.0025$, $p=2.5$, $\gamma_m=30$, $\gamma_{\rm max} = 10^8$ (second panel from the top in \autoref{fig:thPcutaniso}) lights up a spatially extended jet feature in the image with broad and slightly asymmetric (bottom is brighter) emission out to $\sim$1~mas. The asymmetry is triggered by helical bulk motion of the plasma in the jet, which gives rise to stronger relativistic Doppler beaming in the approaching (here bottom) side than the receding (upper) side. However, there is no sign of two-sided limb-brightening in the jet image.

The third panel in \autoref{fig:thPcutaniso} shows the effect of restricting the emission from the jet to a narrow sheath around its outer edge by applying the sigma cutoff prescription in \autoref{eq:msigmacutoff}. Although one might expect that removing emission from the spine region of the jet would contribute to a significantly more limb-brightened and symmetrical jet image, we continue to find a broad and asymmetric jet with no sign of limb-brightening. 
The broad ridge emission in the image comes from the foreground jet sheath, which produces strong isotropic synchrotron emission in a direction perpendicular to the helical magnetic field. 
The asymmetry remains due to a combination of relativistic Doppler beaming and the magnetic field geometry; the helicity of the magnetic fields (caused by frame dragging from the spinning BH) makes the local field lines more aligned with the line of sight on the receding side of the jet, producing weaker emission than on the approaching side.
Thus, we would not observe a symmetric jet image in the isotropic case even without relativistic Doppler beaming.

Finally, as shown in the bottom panel in \autoref{fig:thPcutaniso}, when we introduce anisotropy in the nonthermal electron distribution we do obtain a limb-brightened jet image, similar to what is observed in M87 (compare with the bottom right panel in \autoref{fig:GRFF_obs}). 
Here, synchrotron radiation from a population of nonthermal electrons with strong anisotropy ($\eta=0.01$) produces emission that is concentrated in directions that are aligned with the local magnetic field lines. 
The result is a brightening of both jet edges, where the magnetic field is more closely aligned with the line-of-sight and a suppression of the emission from the foreground sheath region, where the field lines are perpendicular to the line-of-sight. 
In particular, the anisotropy now advantages the de-beamed (upper) side of the jet, in which the magnetic fields are more aligned with the line of sight. 
Notably, this means that the two symmetrical edges of the jet are brightened by two different mechanisms: Doppler beaming on the approaching side, and anisotropic emission aligned with the magnetic field on the receding side.

All the model images shown in the rest of the paper assume anisotropic nonthermal electrons with $\eta=0.01$, and all the GRMHD model images include the sigma cutoff in \autoref{eq:msigmacutoff}\footnote{\edt{An image corresponding to a constant sigma cutoff, $\sigma_m=25$ (a relatively large value suggested by the work of \citealt{Chael24}), is shown in \autoref{apdx:sig25}. It demonstrates that the sigma cutoff contributes very little to the observed limb-brightening in the image.}}.

\subsection{Snapshot Images with the GRMHD Model} \label{subsec:snapshots}

\begin{figure*}
\begin{center}
	\includegraphics[width=18cm]{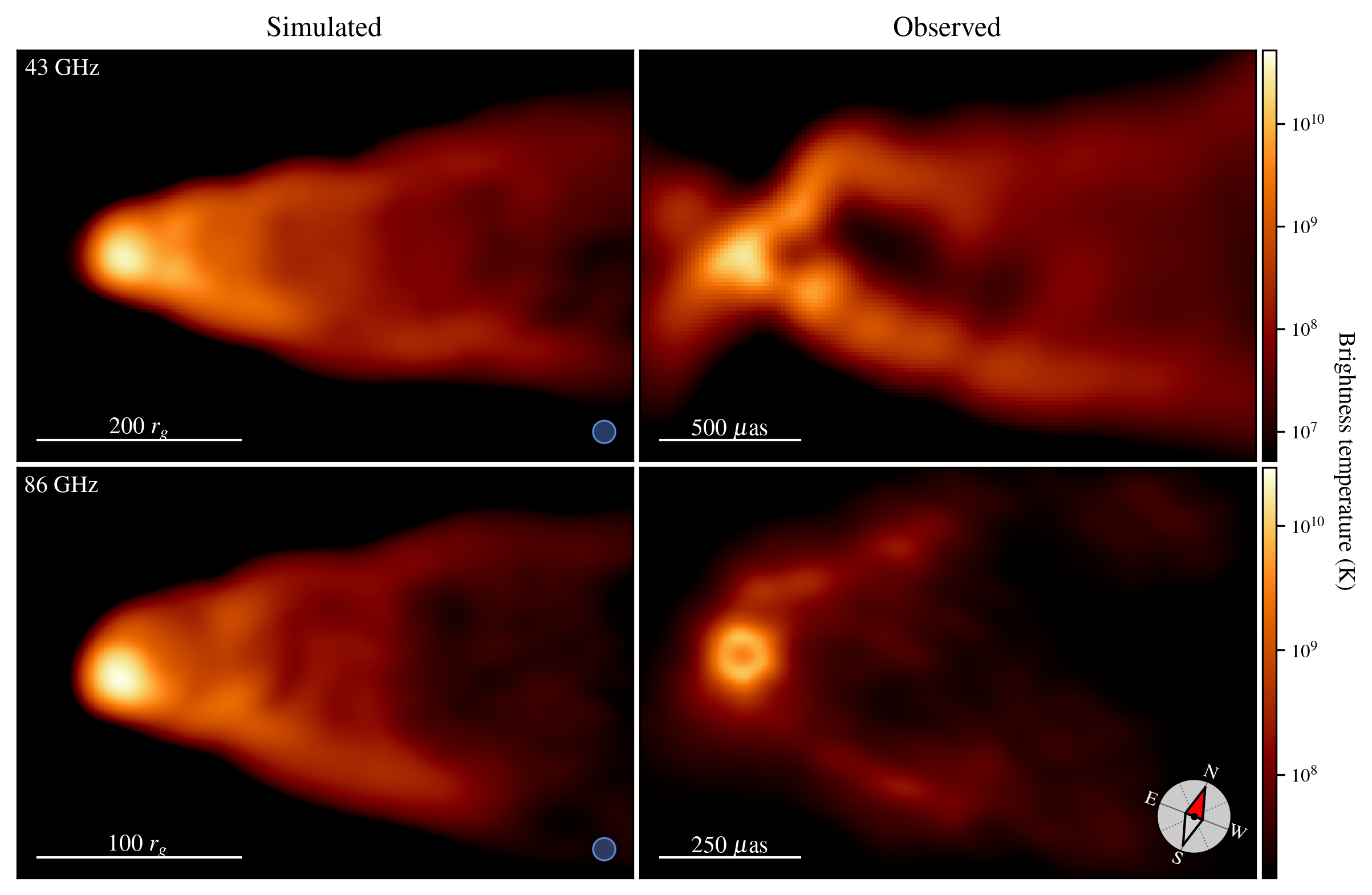}
\end{center}
    \caption{Comparison between  images from a snapshot of the $a_*=0.9$ GRMHD simulation (left column) and single-epoch observed images (right column) of the M87 jet structure at 43\,GHz (top row) and 86\,GHz (bottom row).  The left and right images in each row share a common field of view, with the spatial scale indicated using a scale bar in units of $r_g$ (for the simulated images) or \uas (for the observed images).  We blur the simulated images with circular Gaussian beams of FWHM 40\,\uas and 80\,\uas at 86\,GHz and 43\,GHz, respectively, to approximately match the resolution of the observed images; the FWHM contours for the convolving beams are shown in the lower right-hand corners of the simulated images.
    The 43\,GHz observed image comes from VLBA observations conducted by \citet{2018ApJ...855..128W} and imaged by \citet{2024A&A...690A.129K}, and the 86\,GHz observed image comes from GMVA observations conducted by \citet{2023Natur.616..686L} and imaged by \citet{2024arXiv240900540K}.  The images have been rotated by 21\,degrees so that the jet structure is approximately horizontal; cardinal directions are indicated by the compass on the lower right.
    }
    \label{fig:86snapshot_obs}
\end{figure*}

In \autoref{fig:86snapshot_obs}, we show images at 86 and 43~GHz corresponding to a snapshot from the $a_*=0.9$ GRMHD simulation, and compare them with observations of M87 from \citet{2024arXiv240900540K,2024A&A...690A.129K}. 
The model images are convolved with circular Gaussian beams of $40$ and $80$\,\uas, respectively, to approximately match the angular resolution of the observed images.\footnote{Note that the observational images from \citet{2024arXiv240900540K,2024A&A...690A.129K} were reconstructed using the \texttt{RESOLVE} imaging algorithm \citep{2018arXiv180302174A}, which unlike the more traditional \texttt{CLEAN} algorithm does not explicitly convolve (or ``restore'') the images using a specified Gaussian beam.  Instead, the angular resolution in the reconstructed images is determined by a combination of the data quality and the parameterized image correlation structure, and in general the resolution may even be a function of location within the image.}

The model images show edge-brightening in the jet, though the contrast between the emission at the jet edge and that in the jet center is lower than in the observational images.
We see a number of spiral-shaped components bridging the two edges of the jet; these components originate from the foreground and background sides of the jet sheath, and seemingly analogous features are also seen in the corresponding observational images, though they are more stretched out along the jet. 

In this context, it should be noted that our GRRT calculations are based on the fast light approximation (see \autoref{subsec:GRRT}), under which some of the features in the GRMHD snapshot images are expected to be distorted. Specifically, a slow light calculation will tend to stretch out features longitudinally parallel to the projected jet axis. The amount of stretching will depend on how relativistic the jet motion is, but could be considerable and might well give a better qualitative match between the simulated and observed images.
In future dynamical studies of the M87 jet using data from GRMHD simulations, it will be crucial to use the slow-light ray tracing.

Another noteworthy feature in \autoref{fig:86snapshot_obs} is that the counter-jet component is noticeably fainter in the model images compared to the observational images, a property that also holds for the \grff-based images on larger scales (see \autoref{subsec:GRFF}).
We discuss this issue in \autoref{sec:discussion} in the context of the BH spin-dependence and the prescription for emission anisotropy.

\begin{figure*}
\begin{center}
	\includegraphics[width=18cm]{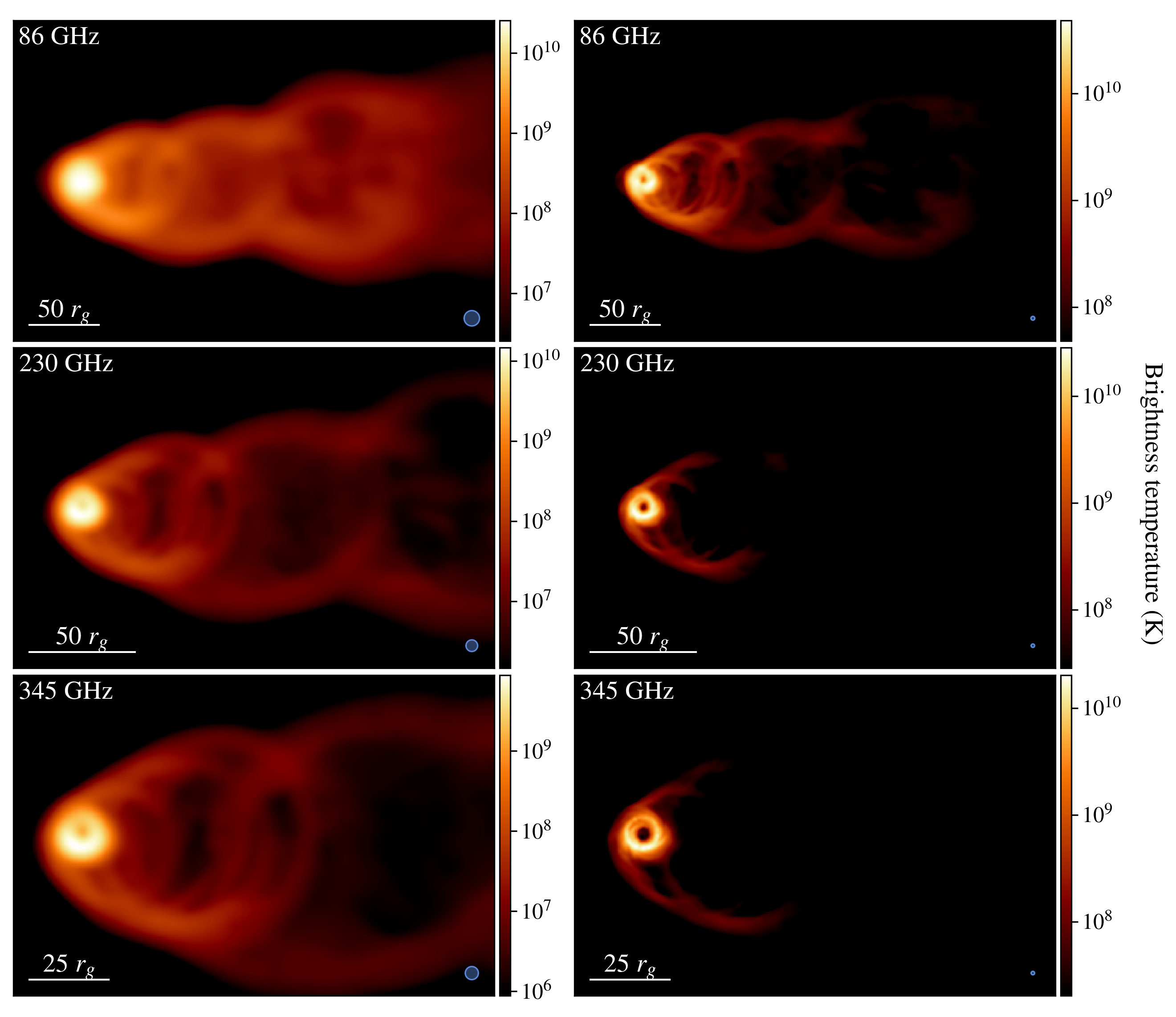}
\end{center}
    \caption{
    Left: Snapshot images from the $a_*=0.9$ GRMHD simulation at 86, 230 and 345~GHz, blurred with circular Gaussian beams of 40, 20, and 15 \uas, respectively, as appropriate for future ngEHT ground-based observations.
    Right: Same as the left, but blurred with smaller beams of 10, 6, and 4 \uas, appropriate for the projected resolution of the future BHEX space VLBI mission. 
    Note that the range of brightness temperatures is different in the left and right columns, and that we use a different snapshot here than the one shown in \autoref{fig:86snapshot_obs}.
    }
    \label{fig:230_345}
\end{figure*}

\autoref{fig:230_345} shows GRMHD snapshot images at the high observing frequencies and fine angular resolutions that are expected to be achievable with next-generation submillimeter VLBI observations with ngEHT \citep{2023Galax..11..107D} and the proposed space-based mission BHEX \citep{2024SPIE13092E..2DJ}. From top to bottom, \autoref{fig:230_345} shows model images at 86, 230, and 345\,GHz blurred to angular resolutions on the left of 40, 20, and 15\,\uas, as appropriate for ngEHT, and angular resolutions on the right of 10, 6, and 4\,\uas, for BHEX.  At these submillimeter wavelengths, plasma in the jet-launching region is optically thin, producing bright ring emission and fainter extended jet emission. 
As a result, we observe arm-like jet components originating from the foreground side of the jet sheath; these components are pronounced in the 230 and 345\,GHz images. 
The arc-like features arise from particularly bright material loaded onto individual field lines, whose propagation traces the magnetic field structure along the jet. Movies made from a sequence of snapshots show the structures moving away from the BH. In this case again slow light GRRT imaging will be essential for quantitative comparisons with observations.

\subsection{Images with the \grff model} \label{subsec:GRFF}

\begin{figure*}
\begin{center}
	\includegraphics[width=18cm]{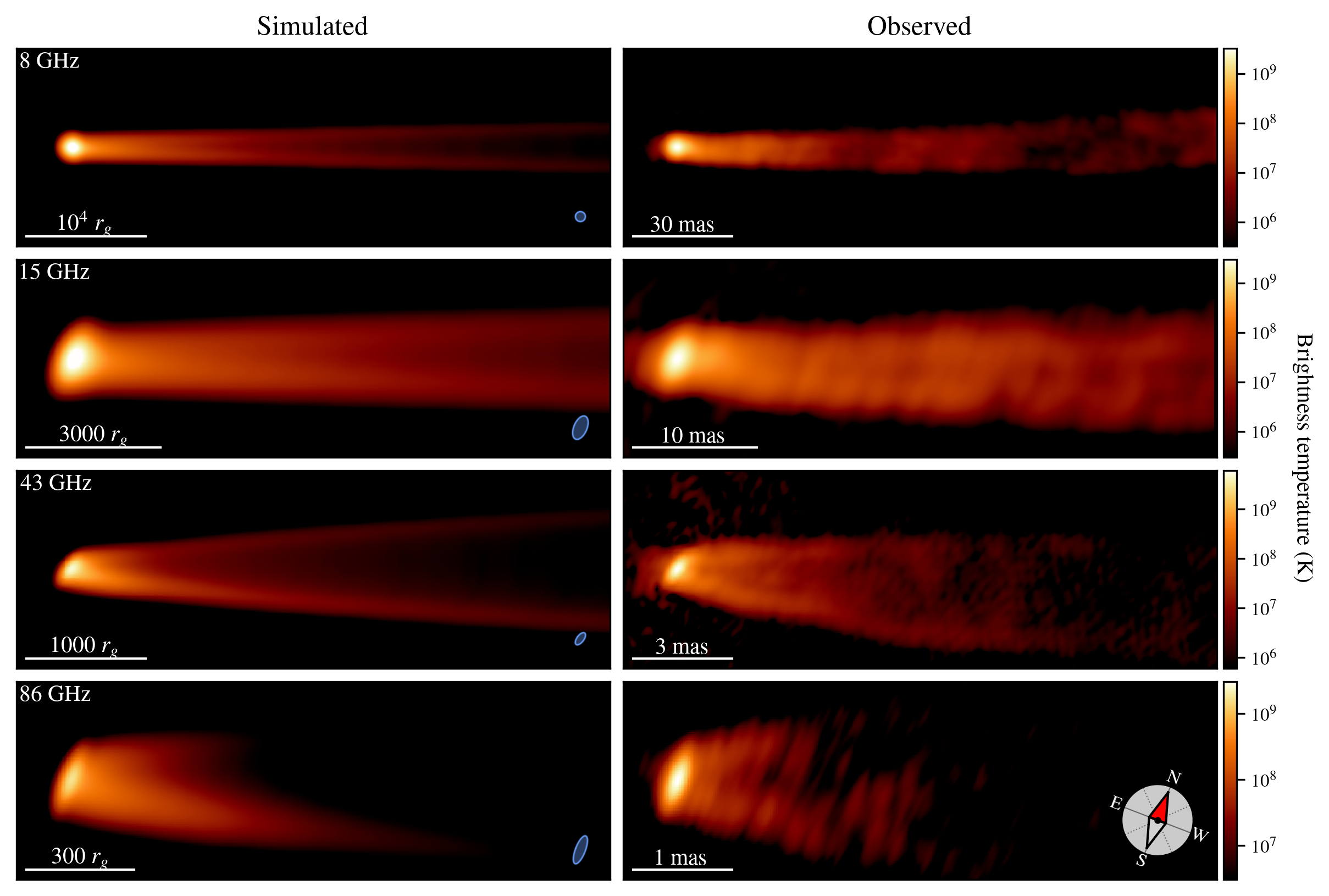}
\end{center}
    \caption{Comparison between simulated \grff images (left column) and observed images (right column) of the M87 jet across four frequencies and at matched spatial scales; from top to bottom, the observing frequencies are 8, 15, 43, and 86\,GHz.  We blur the simulated images using an elliptical Gaussian with the same parameters as the restoring beam used for the corresponding observed images; the FWHM contours for the convolving beams are shown in the lower right-hand corners of the simulated images.  The 8 and 15\,GHz observed images come from single-epoch VLBA observations imaged by \citet{2023MNRAS.526.5949N}, the 43\,GHz observed image comes from multi-epoch time-averaged VLBA observations imaged by \citet{2018ApJ...855..128W}, and the 86\,GHz observed image comes from multi-epoch time-averaged GMVA observations imaged by \citet[][including also HSA observations from \citealt{2016ApJ...817..131H}]{2018A&A...616A.188K}.  All images have been rotated by 21\,degrees so that the jet structure is approximately horizontal; cardinal directions are indicated by the compass on the lower right.
    }
    \label{fig:GRFF_obs}
\end{figure*}

Our GRMHD simulations are only appropriate for spatial scales closer to the BH than $z \lesssim 10^3~r_{\rm g}$, so we switch to the \grff fluid model for describing structure on larger scales.
\autoref{fig:GRFF_obs} shows a comparison between the \grff model images and corresponding M87 observations at 8, 15, 43, and 86\,GHz \citep{2023MNRAS.526.5949N,2018ApJ...855..128W,2018A&A...616A.188K}.
At a glance, the model images show qualitative agreement with the observations in the general shape, brightness, and limb-brightening, and also in the sense and degree of limb (a)symmetry in the approaching jet.

In the 86 and 43~GHz panels of \autoref{fig:GRFF_obs}, the jet shows a weak asymmetry on several-mas scales in which the approaching (bottom) edge is modestly brighter than the receding (upper) edge.
The symmetry apparent at the innermost radii results from a combination of resolution limitations and a balance between the two limb-brightening effects described in \autoref{subsec:average}: Doppler beaming (enhancing emission along the lower edge of the jet) and field-aligned emission (enhancing emission along the upper edge of the jet).
The asymmetry between the two jet limbs that is evident at larger radii in the 86\,GHz image (and to a lesser extent in the 43\,GHz image) arises because at these intermediate radii, the magnetic field is not yet purely toroidal and the velocity field is not yet purely radial (as occurs farther downstream in the jet).  In this particular \grff model, the Doppler beaming that enhances the lower edge of the jet turns out to be moderately more powerful than the anisotropic emission that enhances the upper edge of the jet.  Our model does not produce counter-jet emission that is bright enough to match what is seen in the 43\,GHz observed image.

At larger scales -- as illustrated by the 15 and 8~GHz images in \autoref{fig:GRFF_obs} -- the model produces a symmetrical double-edged jet that extends out to hundreds of milliarcseconds.
In regions of the jet far outside the light cylinder, the bulk plasma velocity is ultra-relativistic with predominantly poloidal motion, while the magnetic fields are almost purely toroidal (\citealp{2009ApJ...697.1164B,2018ApJ...868...82T}). 
This combination leads to strong, symmetrically-beamed emission along both jet limbs and suppressed emission from the central ridge.  We once again find that the counter-jet emission in the model is weaker than that in either the 8 or 15\,GHz observations.

We confirmed that our fiducial model with a ceiling in the bulk velocity of $\gamma_{\rm bulk} = 6$ gives jet images that are consistent with the observations, whereas larger ceiling values cause the jet to appear truncated.\footnote{By contrast, it was confirmed that the suppression factor does not require fine-tuning and is set to a straightforward value of 0.5.} 
This can be understood as follows; 
the relativistic Doppler beaming factor for a purely longitudinal, \edt{approaching} plasma bulk motion and an inclination angle \edt{$i$ $(> 90^\circ)$} can be written as \edt{$g = 1/[\gamma_{\rm bulk} \{1-\beta_{\rm bulk}\,{\rm cos}\,(180^\circ -i)\}]$}, which  for $i = 163^\circ$ peaks at $\gamma_{\rm bulk} \sim 3.5$ and decreases for larger $\gamma_{\rm bulk}$.\footnote{This also implies that we may preferentially observe jet components with $\gamma_{\rm bulk} \sim 3-4$ if there is a differential longitudinal velocity distribution in the outer M87 jet.} 
Thus, if the jet experiences unrestricted acceleration up to $\gamma_{\rm bulk} \gtrsim 10$ as predicted by the analytical \grff model, the observed intensity will decline rapidly in the outer jet and the model would be inconsistent with the extended jet images seen in M87. 
This is confirmed by observations of kinematics in the M87 jet which suggest bulk Lorentz factor values below 10 in the regions within $z \lesssim 10^5~r_{\rm g}$ (\citealp{2019ApJ...887..147P}).

\section{Discussion and Conclusions} \label{sec:discussion}

\begin{figure*}
\begin{center}
    \includegraphics[width=0.8\textwidth]{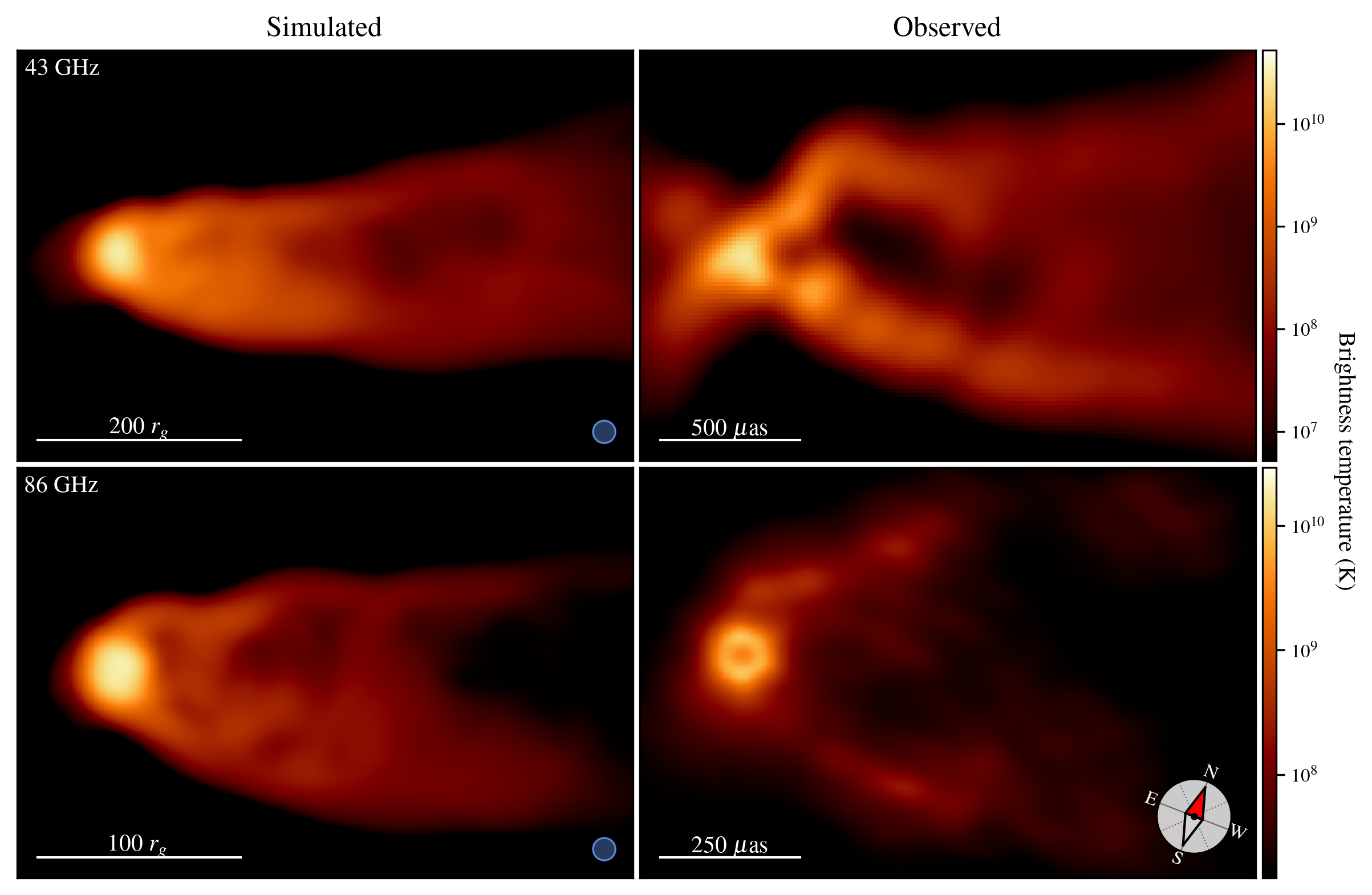}
    \includegraphics[width=1.0\textwidth]{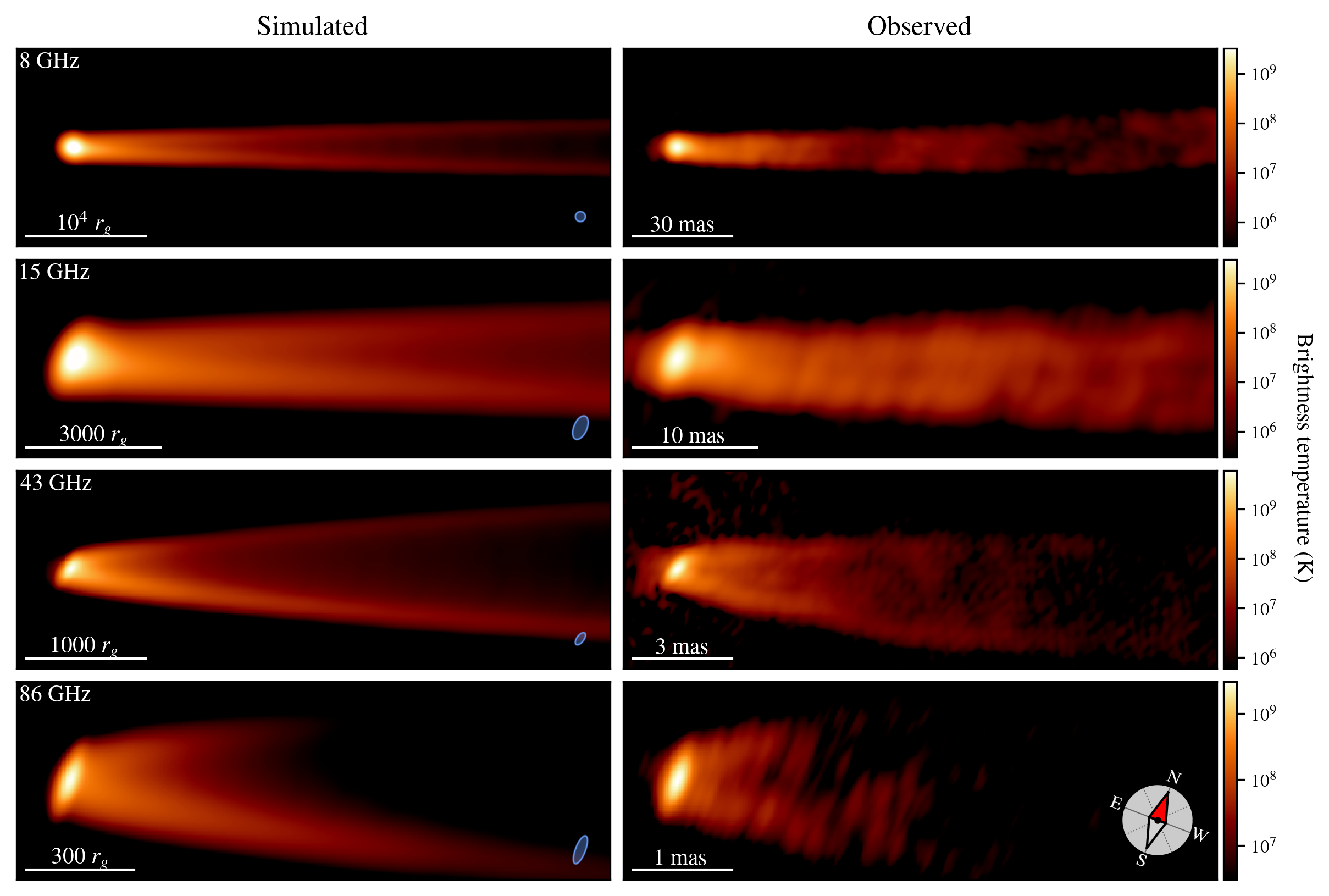}
\end{center}
    \caption{Same as \autoref{fig:86snapshot_obs} (top) and \autoref{fig:GRFF_obs} (bottom) but for the moderate spin ($a_*=0.5$) model.
    }
    \label{fig:GRFF_obs_a05}
\end{figure*}

In this paper we present a model capable of producing the limb brightening seen in observations of astrophysical jets launched from black holes, and we demonstrate that our model can qualitatively and quantitatively reproduce VLBI observations of the jet in the M87 system across more than an order of magnitude in observing frequency ($\sim$8--86\,GHz) and at least three orders of magnitude in spatial scale ($\sim$0.2--200 milliarcseconds).  The radio emission from the jet in our model is synchrotron radiation from a population of nonthermal electrons, which derive their energy from the Poynting power in the jet and subsequently cool (via synchrotron radiation) to produce a number density that follows a double power-law distribution.  

The key new ingredient in our model for producing jet limb-brightening is the inclusion of anisotropy in the electron energy distribution function, such that electrons with velocities parallel to magnetic field lines are strongly favored relative to electrons with perpendicular velocities.  We model the magnetic field and bulk velocity of the plasma throughout the jet using both GRMHD (to describe small-scale and time-resolved structure) and \grff (to describe large-scale and time-averaged structure), and we compute the emission using GRRT with synchrotron emission and absorption coefficients appropriately modified to incorporate the anisotropy. As we demonstrate in Figure~\ref{fig:thPcutaniso}, anisotropy has a large effect on the degree of limb-brightening in the jet, and indeed is essential, within the framework of our model, to produce any limb-brightening at all.

The primary results of this paper are illustrated in Figures \ref{fig:86snapshot_obs}--\ref{fig:GRFF_obs} and can be summarized as follows:
\begin{itemize}
  \item Our GRMHD model enables us to study jet structure and dynamics on spatial scales up to $z \approx 10^3$\,$r_g$.  \autoref{fig:86snapshot_obs} shows that individual snapshots from our GRMHD model can demonstrate qualitative similarity with corresponding observations of the M87 jet system, including the general shape of the jet and the presence of limb brightening.
  \item \autoref{fig:230_345} provides predictions for the appearance of the M87 jet as seen with next-generation ground- and space-based high-frequency VLBI observations.
  \item Our \grff model is applicable only to time-averaged jet structure but can describe arbitrarily large spatial scales.  \autoref{fig:GRFF_obs} shows that images from our \grff model can reproduce the shape and brightness of the M87 jet out to scales of nearly ${\sim}10^5 r_g \approx 250$\,mas.  The presence of limb-brightening and the magnitude of its asymmetry are also qualitatively consistent with observations.
\end{itemize}
\noindent Of the two different values of the BH spin that we explore in this paper, we find that our ``fiducial'' $a_* = 0.9$ model produces images that appear somewhat more consistent with the available M87 observations than those from our $a_* = 0.5$ model, so Figures \ref{fig:86snapshot_obs}--\ref{fig:GRFF_obs} present images from the fiducial $a_* = 0.9$ model.  A comparison between our model images and the observations for the $a_* = 0.5$ case is shown in \autoref{fig:GRFF_obs_a05}.

In this paper we have focused primarily on the ability of anisotropic synchrotron emission to yield limb-brightened jet structures, showing that GRMHD and \grff jet models coupled to strongly ($\eta=0.01$) anisotropic GRRT can produce images that look like observations of M87 across a range of frequencies and spatial scales.  However, we have not comprehensively explored the space of potential model variants, instead adopting simple treatments where possible for a number of the model parameters.  

For instance, we use only a single value of the anisotropy parameter $\eta$ for all nonthermal electrons in our model, rather than coupling the value of $\eta$ to, e.g., plasma-$\beta$ or other physical parameters \citep[as in][]{2023ApJ...957..103G}.  We have not explored the effect of varying the minimum Lorentz factor $\gamma_m$ of the injected nonthermal electrons (see \autoref{subsec:Electrons}). Our treatment of the bulk plasma velocity in the \grff model is also simplistic, with a simple ad hoc scaling factor applied to suppress the velocities to near-GRMHD levels (see \autoref{subsec:GRFFmodel}).  Perhaps because of these limitations, the present version of our model exhibits some points of disagreement with observations; in particular, our model struggles to produce images with a counter-jet as prominent as the one seen in observations\footnote{\edt{All the results in this paper are for an observer inclination angle of $163^\circ$, based on the analysis of \cite{2018ApJ...855..128W}. To investigate if we could make the counterjet substantially brighter by using a less extreme viewing angle, we computed an image for an inclination angle of $155^\circ$, which is outside the range allowed by \cite{2018ApJ...855..128W}. The counterjet remains very dim.}}.  We leave it to future work to explore whether increased model sophistication or physical self-consistency can alleviate such disagreements.

Limitations notwithstanding, we have shown here that a conceptually simple modification to otherwise fairly standard GRMHD or \grff jet models can generically produce the limb-brightening phenomenon that has been so ubiquitous in observations of jetted systems and yet so difficult to reproduce in simulations.  And by virtue of the scale-invariance of GRMHD and \grff, our model is applicable to jetted BH systems beyond just M87 through appropriate choice of the underlying model parameters (BH mass, spin, inclination angle, and accretion rate).  We thus anticipate substantial utility of this model for interpreting existing and future observations of BH jets.

\begin{acknowledgments}
The authors thank Kazunori Akiyama, Yuri Kovalev and Alexander Plavin for constructive discussion and comments, \edt{and the referee for helpful suggestions}. 
We also acknowledge Craig Walker and Jongseo Kim for providing their M87 image reconstructions, and also Alexey Nikonov and Jae-Young Kim for making their own reconstructions available online. 
YT is grateful for support from JSPS (Japan Society for the Promotion of Science) Overseas Research Fellowship. 
\edt{Support for this work was provided by the NSF through grants AST-1935980 and AST-2034306, and by the Gordon and Betty Moore Foundation through grants GBMF5278 and GBMF10423. This work has been supported in part by the Black Hole Initiative at Harvard University, which is funded by the John Templeton Foundation (grants 60477, 61479, and 62286) and the Gordon and Betty Moore Foundation (grant GBMF8273).}

\end{acknowledgments}
%
\vspace{5mm}





\appendix

\section{Electron number density distribution} \label{app:eDF}

In this section we derive the form of the electron number density distribution $f(\gamma)$ -- where $f(\gamma) d\gamma$ provides the number density of electrons with Lorentz factors in the interval $d\gamma$ -- given the prescription described in \autoref{subsec:Electrons}.  We assume that the electrons are initially injected with a power-law distribution of Lorentz factors,
\begin{equation}
f_{\text{inj}}(\gamma) = 
\begin{dcases}
0 , & \gamma < \gamma_m \\
n_m \left(\frac{\gamma}{\gamma_m}\right)^{-p} , & \gamma_m \leq \gamma \leq \gamma_{\text{max}} \\
0 , & \gamma > \gamma_{\text{max}}
\end{dcases} , \label{eqn:InjectedElectronDensity}
\end{equation}
\noindent where $\gamma_m$ and $\gamma_{\text{max}}$ are the minimum and maximum Lorentz factors of the injected power-law distribution, respectively, 
$n_m$ is the number density of electrons injected with Lorentz factor $\gamma_m$,
and $p$ is the power-law index. 
Here, $n_m = n_{\rm pl} (p-1)/\{\gamma_m ^p(\gamma_m ^{1-p}-\gamma_{\rm max}^{1-p})\}$ and $n_{\rm pl}$ is the total number density determined from the Poynting flux as described in \autoref{subsec:Electrons}. We further assume that the electrons cool via synchrotron radiation.  Similar derivations can be found in, e.g., \cite{1970ranp.book.....P}, \cite{2011A&A...529A..47M}, and \cite{2019ApJ...886..106P}.

\subsection{Single Injection Event}

The total power radiated by an electron with Lorentz factor $\gamma$ in a magnetic field of strength $B$ is given by
\begin{equation}
\dot{E} = -\frac{\sigma_T c \gamma^2 B^2}{6 \pi} .
\end{equation}
\noindent Given an energy per electron of $E = \gamma m_e c^2$, the corresponding rate of change of the electron's Lorentz factor will be\footnote{Equations~\ref{Eq:gammadot} and \ref{Eq:Gamma} are standard results for an isotropic energy distribution function. A more careful analysis, which is left to future work, will need to include the effects of anisotropy.}
\begin{equation}
\dot{\gamma} = \frac{\dot{E}}{m_e c^2} = - \frac{\sigma_T \gamma^2 B^2}{6 \pi m_e c} .
\label{Eq:gammadot}
\end{equation}
\noindent Assuming that $B$ is constant in time, integrating $\dot{\gamma}$ yields
\begin{equation}
\gamma(t) = \left( \frac{1}{\gamma_0} + \frac{1}{\Gamma(t)} \right)^{-1} ,
\end{equation}
\noindent where $\gamma_0$ is the Lorentz factor at $t=0$ and 
\begin{equation}
\Gamma(t) \equiv \frac{6 \pi m_e c}{\sigma_T B^2 t} .
\label{Eq:Gamma}
\end{equation}
\noindent By continuity of electron number, we have for a single injection event
\begin{equation}
f_s(\gamma,t) = f_{\text{inj}}(\gamma_0) \frac{d\gamma_0}{d\gamma} ,
\end{equation}
\noindent which evaluates to 
\begin{equation}
f_s(\gamma,t) = \begin{dcases}
0 , & \gamma < \left( \frac{1}{\gamma_m} + \frac{1}{\Gamma(t)} \right)^{-1} \\
n_m \left( \frac{\gamma}{\gamma_m} \right)^{-p} \left( 1 - \frac{\gamma}{\Gamma(t)} \right)^{p-2} , & \left( \frac{1}{\gamma_m} + \frac{1}{\Gamma(t)} \right)^{-1} \leq \gamma \leq \left( \frac{1}{\gamma_{\text{max}}} + \frac{1}{\Gamma(t)} \right)^{-1} \\
0 , & \gamma > \left( \frac{1}{\gamma_{\text{max}}} + \frac{1}{\Gamma(t)} \right)^{-1} \\
\end{dcases} . 
\end{equation}
Here, $n_m$ is the number density of electrons injected with Lorentz factor $\gamma_m$.

\subsection{Continuous Injection}

\begin{figure}[t]
    \centering
    \includegraphics[width=0.50\textwidth]{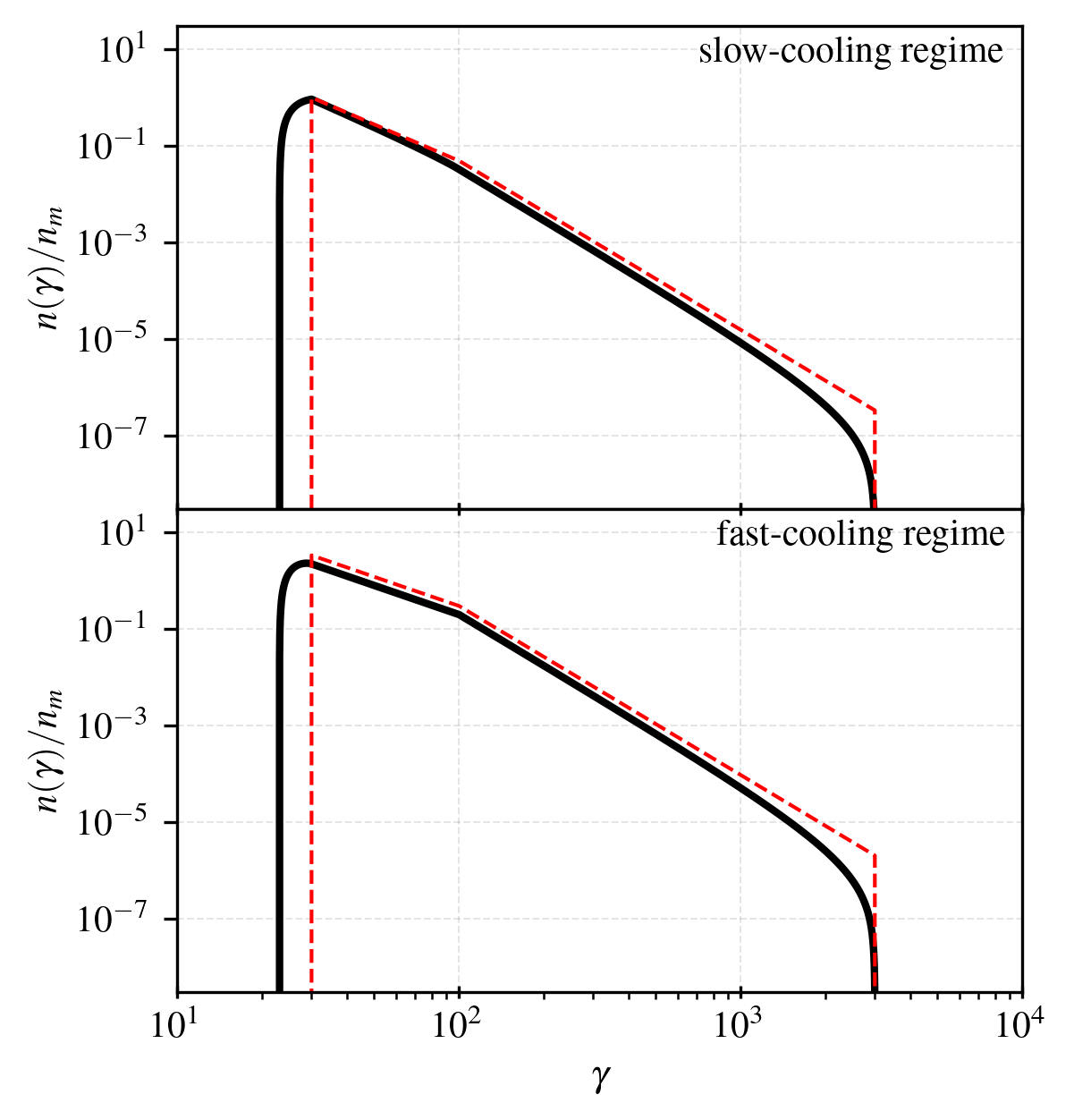}
    \caption{Example electron energy distribution functions corresponding to the slow-cooling (upper panel) and fast-cooling (lower panel) regimes.  The solid black curves show the exact distributions from \autoref{eqn:SlowCoolingExact} and \autoref{eqn:FastCoolingExact}, and the red dashed curves show the double power-law approximations from \autoref{eqn:SlowCoolingApprox} and \autoref{eqn:FastCoolingApprox}.  For the upper panel we have set $\gamma_m = 30$ and $\gamma_c =100$, while the lower panel uses $\gamma_m = 100$ and $\gamma_c = 30$; we have set $\gamma_{\text{max}} = 3000$ for both panels.}\label{fig:DoublePLelectrons}
\end{figure}

However, the total electron population at some time $t_c$ will actually contain contributions from electron sub-populations that have had cooling times ranging from $t=0$ up to $t = t_c$, so we need to integrate further.  The integral setup looks like
\begin{equation}
f(\gamma) = \frac{\gamma_c}{t_c} \int_{0}^{t_c} f_s(\gamma, t_c-t) dt\,,
\end{equation}
\noindent where $\gamma_c \equiv \Gamma(t_c)$ and the $\gamma_c/t_c$ prefactor ensures that the distribution remains normalized to the injected value.  Evaluating this integral requires tracking several possible branches determined by the relative values of various Lorentz factors.  Defining
\begin{equation}
\gamma_{\text{min}} \equiv \left( \frac{1}{\gamma_m} + \frac{1}{\gamma_c} \right)^{-1}
\end{equation}
\noindent and
\begin{equation}
\gamma_{\text{mid}} \equiv \left( \frac{1}{\gamma_{\text{max}}} + \frac{1}{\gamma_c} \right)^{-1} ,
\end{equation}
\noindent the behavior of $n(\gamma)$ depends on the relative values of $\gamma_{\text{mid}}$ and $\gamma_m$.  If $\gamma_m < \gamma_{\text{mid}}$, then we are in the ``slow-cooling regime'' and
\begin{equation}
f_{\text{slow}}(\gamma) = \begin{dcases}
0 , & \gamma < \gamma_{\text{min}} \\
\frac{n_m \gamma_c}{(p-1) \gamma} \left( \frac{\gamma}{\gamma_m} \right)^{-p} \left[ \left( \frac{\gamma}{\gamma_m} \right)^{p-1} - \left( 1 - \frac{\gamma}{\gamma_c} \right)^{p-1} \right] , & \gamma_{\text{min}} \leq \gamma < \gamma_m \\
\frac{n_m \gamma_c}{(p-1) \gamma} \left( \frac{\gamma}{\gamma_m} \right)^{-p} \left[ 1 - \left( 1 - \frac{\gamma}{\gamma_c} \right)^{p-1} \right] , & \gamma_m \leq \gamma < \gamma_{\text{mid}} \\
\frac{n_m \gamma_c}{(p-1) \gamma} \left( \frac{\gamma}{\gamma_m} \right)^{-p} \left[ 1 - \left( \frac{\gamma}{\gamma_{\text{max}}} \right)^{p-1} \right] , & \gamma_{\text{mid}} \leq \gamma < \gamma_{\text{max}} \\
0 , & \gamma \geq \gamma_{\text{max}}
\end{dcases} . \label{eqn:SlowCoolingExact}
\end{equation}
\noindent Similarly, if $\gamma_m > \gamma_{\text{mid}}$, then we are in the ``fast-cooling regime'' and
\begin{equation}
f_{\text{fast}}(\gamma) = \begin{dcases}
0 , & \gamma < \gamma_{\text{min}} \\
\frac{n_m \gamma_c}{(p-1) \gamma} \left( \frac{\gamma}{\gamma_m} \right)^{-p} \left[ \left( \frac{\gamma}{\gamma_m} \right)^{p-1} - \left( 1 - \frac{\gamma}{\gamma_c} \right)^{p-1} \right] , & \gamma_{\text{min}} \leq \gamma < \gamma_{\text{mid}} \\
\frac{n_m \gamma_c}{(p-1) \gamma} \left( \frac{\gamma}{\gamma_m} \right)^{-1} \left[ 1 - \left( \frac{\gamma_m}{\gamma_{\text{max}}} \right)^{p-1} \right] , & \gamma_{\text{mid}} \leq \gamma < \gamma_m \\
\frac{n_m \gamma_c}{(p-1) \gamma} \left( \frac{\gamma}{\gamma_m} \right)^{-p} \left[ 1 - \left( \frac{\gamma}{\gamma_{\text{max}}} \right)^{p-1} \right] , & \gamma_m \leq \gamma < \gamma_{\text{max}} \\
0 , & \gamma \geq \gamma_{\text{max}}
\end{dcases} . \label{eqn:FastCoolingExact}
\end{equation}
\noindent The distributions \autoref{eqn:SlowCoolingExact} and \autoref{eqn:FastCoolingExact} are shown as the solid black curves in \autoref{fig:DoublePLelectrons}.

\subsection{Double Power-Law Approximation}

The distribution functions \autoref{eqn:SlowCoolingExact} and \autoref{eqn:FastCoolingExact} can be approximated as double power-laws \citep[e.g.,][]{1998ApJ...497L..17S}, which simplifies computation of the synchrotron emissivity and absorption coefficients needed for downstream GRRT.  The distribution function in the slow-cooling regime ($\gamma_c > \gamma_m$) can be approximated as
\begin{eqnarray}
f_{\text{slow}}(\gamma) \approx \begin{dcases}
0 , & \gamma < \gamma_m \\
n_m \left( \frac{\gamma}{\gamma_m} \right)^{-p} , & \gamma_m \leq \gamma < \gamma_c \\
n_m \left( \frac{\gamma_c}{\gamma_m} \right)^{-p} \left( \frac{\gamma}{\gamma_c} \right)^{-(p+1)} , & \gamma_c \leq \gamma < \gamma_{\text{max}} \\
0 , & \gamma \geq \gamma_{\text{max}}
\end{dcases} , \label{eqn:SlowCoolingApprox}
\end{eqnarray}
\noindent while in the fast-cooling regime ($\gamma_c < \gamma_m$) we have
\begin{eqnarray}
f_{\text{fast}}(\gamma) \approx \begin{dcases}
0 , & \gamma < \gamma_c \\
n_m \left( \frac{\gamma_m}{\gamma_c} \right) \left( \frac{\gamma}{\gamma_c} \right)^{-2} , & \gamma_c \leq \gamma < \gamma_m \\
n_m \left( \frac{\gamma_c}{\gamma_m} \right) \left( \frac{\gamma}{\gamma_m} \right)^{-(p+1)} , & \gamma_m \leq \gamma < \gamma_{\text{max}} \\
0 , & \gamma \geq \gamma_{\text{max}}
\end{dcases} . \label{eqn:FastCoolingApprox}
\end{eqnarray}
\noindent The distributions \autoref{eqn:SlowCoolingApprox} and \autoref{eqn:FastCoolingApprox} are shown as the dashed red curves in \autoref{fig:DoublePLelectrons}.



\bigskip
\section{Anisotropic double power-law synchrotron emissivity and absorption coefficient} \label{apdx:aniso_wPL}


To generalize the anisotropic single power-law model in Eq.~\ref{eq:anisopl} to an anisotropic double power-law distribution function with a lower cutoff at $\gamma=\gamma_{\rm min}$ and a break at $\gamma = \gamma_{\rm br}$, we write 
\begin{eqnarray} \label{eq:wpl_aniso}
  f(\gamma,\xi) = \begin{dcases}
      N_{\rm wpl} \ \gamma^{-p_1} \left\{1 + (\eta-1){\rm cos}^2\xi \right\}^{-p_1/2} & (\gamma_{\rm min} < \gamma < \gamma_{\rm br}, \ 0 \leq \xi \leq \pi) \\
      N_{\rm wpl} \ \frac{P(p_1,\eta)}{P(p_2,\eta)} {\gamma_{\rm br}}^{(p_2 - p_1)} \ \gamma^{-p_2} \left\{1 + (\eta-1){\rm cos}^2\xi \right\}^{-p_2/2} & (\gamma_{\rm br} < \gamma < \gamma_{\rm max}, \ 0 \leq \xi \leq \pi) \\
      0 \ & ( \, {\rm otherwise} \, ). 
  \end{dcases}   
\end{eqnarray}
Here, the two power-law parts are connected at 
$\gamma = \gamma_{\rm br}$ after integrating over $\xi$, 
and $N_{\rm wpl}$ is a normalization factor. Note that, for slow cooling, $\gamma_{\rm min}=\gamma_{\rm m}$, $p_1=p$, $\gamma_{\rm br}=\gamma_c$, $p_2=p+1$, whereas for fast cooling,  $\gamma_{\rm min}=\gamma_{\rm c}$, $p_1=2$, $\gamma_{\rm br}=\gamma_m$, $p_2=p+1$.

Normalizing $f(\gamma,\xi)$ by the total number density of power-law electrons $n_{\rm pl}$ yields
\begin{eqnarray} \label{eq:wpl_aniso_norm}
    N_{\rm wpl} &=& \frac{(p_1-1)n_{\rm pl}}{A} P(p_1,\eta)^{-1}, \\
    A &=& \gamma_{\rm min}^{1-p_1}-\gamma_{\rm br}^{1-p_1} + \frac{p_1-1}{p_2-1} \gamma_{\rm br}^{p_2-p_1} (\gamma_{\rm br}^{1-p_2}-\gamma_{\rm max}^{1-p_2}) . 
\end{eqnarray}
For $\eta=1$, Eq.~\ref{eq:wpl_aniso} reverts to an isotropic double power-law distribution function (see, for example, \citealp{2023ApJ...949..101K}).
As the two power-law segments, $\gamma_{\rm min} < \gamma < \gamma_{\rm br}$ and $\gamma_{\rm br} < \gamma < \gamma_{\rm max}$, in Eq.~\ref{eq:wpl_aniso} contribute to the two corresponding segments of the $\gamma$-integral and are then summed up, the full-polarization, anisotropic double power-law synchrotron emissivities is obtained as 
\begin{eqnarray} \label{eq:wpl_aniso_j}
    j_I &=& \phi_1(\theta_B) j_{I,{\rm iso1}} + \gamma_{\rm br}^{p_2-p_1} \phi_2(\theta_B) j_{I,{\rm iso2}}, \\
    j_Q &=& \phi_1(\theta_B) j_{Q,{\rm iso1}} + \gamma_{\rm br}^{p_2-p_1} \phi_2(\theta_B) j_{Q,{\rm iso2}}, \\
    j_U &=& 0, \\
    j_V &=& \phi_1(\theta_B) \left\{ 1 + \frac{g_1(\theta_B)}{p_1+2} \right\} j_{V,{\rm iso1}} + \gamma_{\rm br}^{p_2-p_1} \phi_2(\theta_B) \left\{ 1 + \frac{g_2(\theta_B)}{p_2+2} \right\} j_{V,{\rm iso2}}, 
\end{eqnarray}
where
\begin{eqnarray}
    \phi_1(\theta_B) = P(p_1,\eta)^{-1} \left\{1 + (\eta-1){\rm cos}^2\theta_B \right\}^{-p_1/2}, \ &&\phi_2(\theta_B) = P(p_2,\eta)^{-1} \left\{1 + (\eta-1){\rm cos}^2\theta_B \right\}^{-p_2/2}, \\
    g_1(\theta_B) = \frac{p_1(\eta-1){\rm sin}^2\theta_B}{1 + (\eta-1){\rm cos}^2\theta_B}, \ &&g_2(\theta_B) = \frac{p_2(\eta-1){\rm sin}^2\theta_B}{1 + (\eta-1){\rm cos}^2\theta_B}.
\end{eqnarray}
Here, $j_{\{I,Q,V\},{\rm iso1}}$ and $j_{\{I,Q,V\},{\rm iso2}}$ are the isotropic single power-law emissivities for $\gamma_{\rm min} < \gamma < \gamma_{\rm br}$ with power-law index $p_1$ and $\gamma_{\rm br} < \gamma < \gamma_{\rm max}$ with $p_2$, respectively, with the total number density of power-law electron given by $n_{\rm pl}$. 
The absorption coefficients are also obtained in an analogous way as follows; 
\begin{eqnarray}
    \alpha_I &=& \phi_1(\theta_B) \alpha_{I,{\rm iso1}} + \gamma_{\rm br}^{p_2-p_1} \phi_2(\theta_B) \alpha_{I,{\rm iso2}}, \\
    \alpha_Q &=& \phi_1(\theta_B) \alpha_{Q,{\rm iso1}} + \gamma_{\rm br}^{p_2-p_1} \phi_2(\theta_B) \alpha_{Q,{\rm iso2}}, \\
    \alpha_U &=& 0, \\
    \alpha_V &=& \phi_1(\theta_B) \left\{ 1 + \frac{g_1(\theta_B)}{p_1+2} \right\} \alpha_{V,{\rm iso1}} + \gamma_{\rm br}^{p_2-p_1} \phi_2(\theta_B) \left\{ 1 + \frac{g_2(\theta_B)}{p_2+2} \right\} \alpha_{V,{\rm iso2}}.
\end{eqnarray}

\bigskip
\section{\edt{Poynting Flux}} \label{apdx:poynting}
\edt{In our model, we set the non-thermal electron energy density by requiring that it be proportional to the magnitude of the Poynting vector $\vec{S}$ (see Eq.~\ref{eq:uplinj}). However, $|\vec{S}|$ is a frame-dependent quantity, so we must choose a frame in which to perform this density normalization. In the Kerr spacetime, the natural candidate is the Zero-Angular-Momentum (ZAMO) frame, which remains timelike everywhere in the black hole exterior and is given in Boyer-Lindquist coordinates }by\begin{align}
    n_\mu&=(-\alpha,0,0,0)
\end{align}
\edt{with $\alpha=(-g^{tt})^{-1/2}$.  }

\edt{The Poynting flux 4-vector in the ZAMO frame is given by an appropriate projection of the stress-energy, $T^{\mu\nu}_{\rm EM}$:}
\begin{align}
    S^\mu&= \left(n^\mu n_\nu + \delta^{\mu}_{\nu}\right)\left(-n_\alpha T^{\alpha\nu}_{\rm EM}\right).    
\end{align}
\edt{This definition of $S^\mu$ is equivalent to the electromagnetic part of the energy flux vector perpendicular to the normal observer, $\tilde{Q}^\mu$ in \citet{noble2006primitive}. The magnitude of the normal observer Poynting flux is:}
\begin{equation}
\label{eq:Smag}
    |\vec{S}|_{\rm ZAMO}=\sqrt{S^\mu S_\mu}=\sqrt{v_\perp^2 \mathcal{B}^4},
\end{equation}
\edt{where $v_\perp$ is the field-perpendicular fluid velocity as measured in the ZAMO frame, and $\mathcal{B}$ is the magnetic field strength as measured in the ZAMO frame. Both quantities can be defined covariantly as}
\begin{align}
\label{eq:vperp}
  v_\perp^\mu&=\frac{u^\mu}{\gamma}-v_\parallel\frac{\mathcal{B}^\mu}{\sqrt{\mathcal{B}^2}}-n^\mu,\qquad  \mathcal{B}^\mu=n_\nu (\star F)^{\nu\mu},
\end{align}
\edt{with $u^\mu$ the plasma four-velocity, $\star F$ the Faraday dual, $\gamma=-u_\mu n^\mu$ the ZAMO frame Lorentz factor, and $v_\parallel=u_\mu \mathcal{B}^\mu / \gamma\sqrt{\mathcal{B}^2}$ the ZAMO frame field-parallel velocity. Furthermore, one can show \citep[e.g.][Eq. 24]{Chael24} that $v^2_\perp = \mathcal{E}^2/\mathcal{B}^2$, where $\mathcal{E}^\mu=n_\nu F^{\mu\nu}$ is the ZAMO frame electric field, so that}
\begin{equation}
\label{eq:Smag2}
    |\vec{S}|_{\rm ZAMO}=\sqrt{\mathcal{E}^2 \mathcal{B}^2}.
\end{equation}


\edt{One can alternatively derive Eqs~\ref{eq:Smag},~\ref{eq:Smag2} by explicitly constructing an orthonormal tetrad for the ZAMO frame. Since this orthonormal frame is locally Minkowski, one can use $\vec{S}=\frac{c}{4\pi}\vec{E}\times\vec{B}$, with $\vec{E}$ and $\vec{B}$ the electromagnetic three-vectors computed from the tetrad. Then, upon successive application of vector identities combined with the definition of the drift velocity, one can simplify the cross product to arrive at Eq.~\ref{eq:Smag}. }

\bigskip
\section{\edt{Role of the sigma cutoff}} \label{apdx:sig25}

\begin{figure*}
\begin{center}
	\includegraphics[width=14cm]{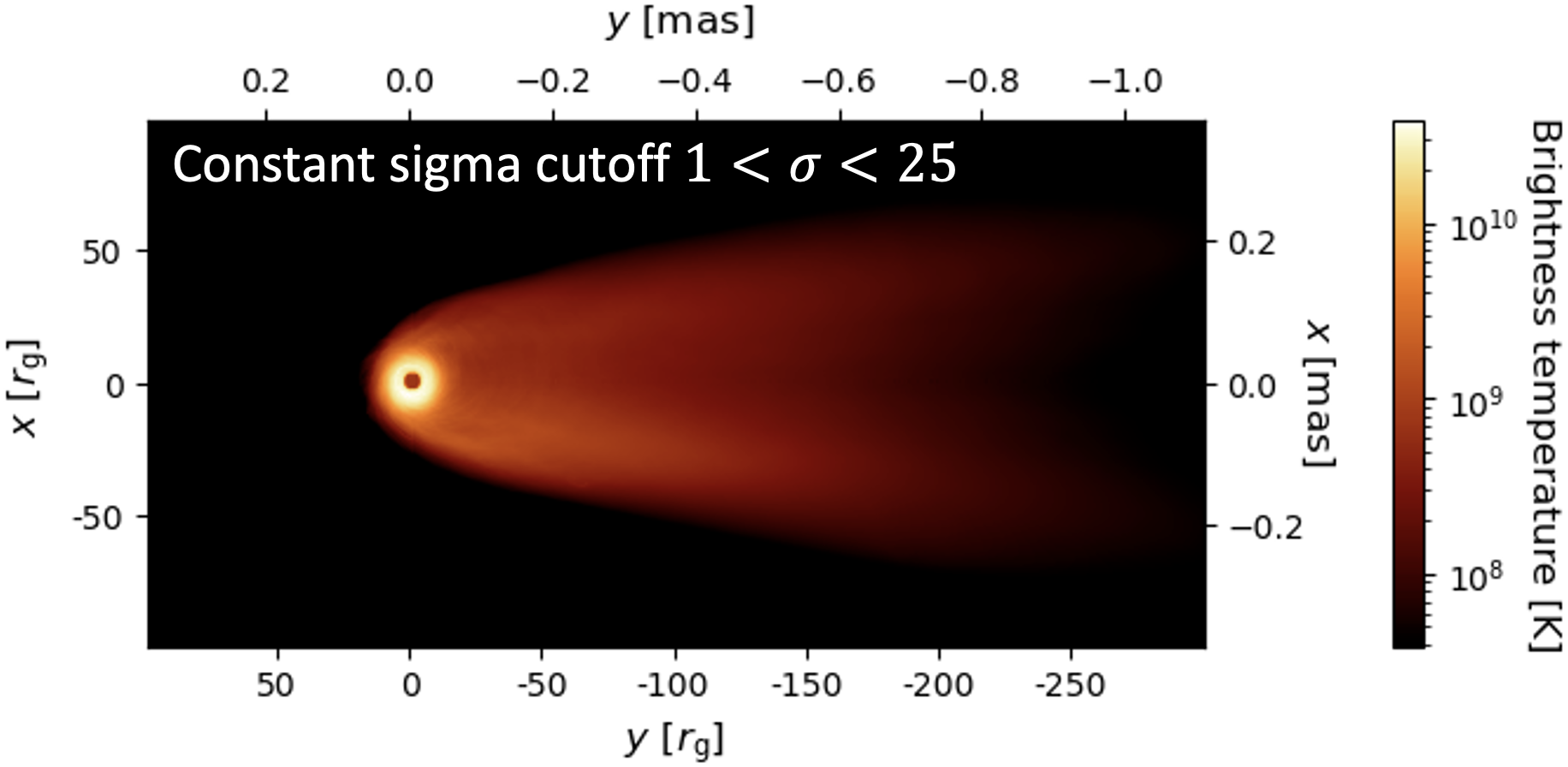}
\end{center}
    \caption{\edt{Same as the bottom panel of \autoref{fig:thPcutaniso}, but with a constant sigma cutoff for non thermal emission, $1 < \sigma < 25$ (based on \citealt{Chael24}). }
    }
    \label{fig:sig25}
\end{figure*}

\edt{In the third and fourth panels in \autoref{fig:thPcutaniso}, we have applied the sigma cutoff prescription given in \autoref{eq:msigmacutoff} in order to limit the region of the GRMHD model that is included when computing the image. Applying a sigma cutoff is very common because the high-sigma regions of a GRMHD simulation are prone to large numerical errors. The very low density in these regions causes numerical difficulties which require the application of density floors and other numerical fixes to stabilize the simulation. This is of course at the cost of accuracy. In much of the work of the EHT Collaboration, a severe cutoff of $\sigma_m=1$ is applied \citep[e.g.,][]{2019ApJ...875L...3E}. Although this cuts out a large part of the jet volume, this is of little consequence for studies that focus on the BH shadow. }

\edt{For studies of the jet, however, it is necessary to relax the sigma cutoff criterion, and there has been some previous discussion \citep[e.g.,][]{2019MNRAS.486.2873C}. In an important recent study, \citet{Chael24}
carried out hybrid GRMHD+GRFFE simulations which treat the jet funnel region more accurately using force-free equations. By comparing the results with those from traditional GRMHD simulations, he finds that GRMHD simulation results may be trusted for $\sigma$ values up to $\sigma_m=25$, but regions with larger $\sigma$ are problematic\footnote{\edt{An additional consideration is that most GRMHD simulations apply so-called reflecting boundary conditions at the poles, which means that the first couple of cells around $\theta=0$ and $\pi$ must be ignored. A sigma cutoff naturally eliminates this region.}}.
Our cutoff prescription (\autoref{eq:msigmacutoff}) is more stringent than Chael's limit for projected distances $y> 40r_g$ in the jet. A natural question then is: How much of the limb-brightening we obtained in the bottom panel in \autoref{fig:thPcutaniso} is from our severe sigma cutoff \autoref{eq:msigmacutoff} rather from our introduction of anisotropy in the electrons.}


\edt{To answer this question, we show in \autoref{fig:sig25} the time-averaged image we obtain when we use \citet{Chael24}'s constant sigma cutoff of $\sigma_m = 25$ for calculating the nonthermal radiation from the jet.
We confirm that, even with this less severe sigma cutoff the image still gives a limb-brightened jet, which looks essentially the same as in the fiducial model image in the bottom panel in \autoref{fig:thPcutaniso}. Thus the limb-brightening is primarily from introducing ansitropy in the electron distribution.
A more detailed comparison reveals that the image in \autoref{fig:sig25} is dimmer near the inner ring and brighter in the outer jet compared to the fiducial model, since the $\sigma_m=25$ cutoff excludes a larger volume of the jet near the BH but a narrower region in the outer funnel, relative to \autoref{eq:msigmacutoff}.}


\bibliography{sample631}{}
\bibliographystyle{aasjournal}



\end{document}